\documentclass{cupconf}
\usepackage{epsf}

  \checkfont{eurm10}
  \iffontfound
    \IfFileExists{upmath.sty}
      {\typeout{^^JFound AMS Euler Roman fonts on the system,
                   using the 'upmath' package.^^J}%
       \usepackage{upmath}}
      {\typeout{^^JFound AMS Euler Roman fonts on the system, but you
                   dont seem to have the}%
       \typeout{'upmath' package installed. cupconf.cls can take advantage
                 of these fonts,^^Jif you use 'upmath' package.^^J}%
      }
  \else
  \fi

  \checkfont{msam10}
  \iffontfound
    \IfFileExists{amssymb.sty}
      {\typeout{^^JFound AMS Symbol fonts on the system, using the
                'amssymb' package.^^J}%
       \usepackage{amssymb}%
       \let\le=\leqslant  
       \let\ge=\geqslant  
      }{}
  \fi

  \IfFileExists{amsbsy.sty}
    {\typeout{^^JFound the 'amsbsy' package on the system, using it.^^J}%
     \usepackage{amsbsy}}
    {\providecommand\boldsymbol[1]{\mbox{\boldmath $##1$}}}

\newsavebox{\astrutbox}
\sbox{\astrutbox}{\rule[-5pt]{0pt}{20pt}}

\def\gsim{\mathrel{\raise0.35ex\hbox{$\scriptstyle >$}\kern-0.6em % Greater/squiggles
\lower0.40ex\hbox{{$\scriptstyle \sim$}}}}
\def\lsim{\mathrel{\raise0.35ex\hbox{$\scriptstyle <$}\kern-0.6em % Less than/squiggles
\lower0.40ex\hbox{{$\scriptstyle \sim$}}}}

\title[Galaxy Assembly]{Galaxy Assembly}

\author[E.\ F.\ Bell]%
{Eric F.\ Bell}

\affiliation{Max-Planck-Institut f\"ur Astronomie, K\"onigstuhl 17, 
D-69117 Heidelberg, Germany \hspace{1.3cm} \texttt{bell@mpia.de}}

\pubyear{}
\volume{}
\pagerange{}
\date{}

\begin{document}

\maketitle

\begin{abstract}
In a $\Lambda$CDM Universe, galaxies grow in mass both through
star formation and through addition of already-formed stars
in galaxy mergers.  Because of this partial decoupling of 
these two modes of galaxy growth, I discuss each separately in this
biased and incomplete review of galaxy assembly; first giving an 
overview of the cosmic-averaged star
formation history, and then moving on to discuss the importance
of major mergers in shaping the properties of present-day 
massive galaxies.  The cosmic-averaged star formation rate, when 
integrated, is in reasonable agreement with the
build-up of stellar mass density.  Roughly
2/3 of all stellar mass is formed during an epoch of rapid star 
formation prior to $z \sim 1$, with
the remaining 1/3 formed in the subsequent 9 Gyr during 
a period of rapidly-declining star formation rate.
The epoch of important star formation in massive galaxies is
essentially over.  In contrast, a significant fraction 
of massive galaxies undergo a major merger at $z \lsim 1$, 
as evidenced by close pair statistics, morphologically-disturbed
galaxy counts, and the build-up of stellar mass in morphologically
early-type galaxies.   Each of these methods is highly uncertain;
yet, taken together, it is not implausible that the massive galaxy population 
is strongly affected by late galaxy 
mergers, in excellent qualitative agreement with 
our understanding of galaxy evolution in a $\Lambda$CDM Universe.
\end{abstract}

\firstsection % if your document starts with a section,
              % remove some space above using this command.
\section{Introduction}

The last decade has witnessed amazing progress in our 
empirical and theoretical understanding of galaxy formation and evolution.  
This explosion in our understanding has been driven
largely by technology: the profusion of 8--10-m class
telescopes, the advent of wide-field
multi-object spectrographs and imagers on large telescopes,
servicing missions for the {\it Hubble Space Telescope} (HST), giving it 
higher resolution, higher sensitivity, larger field-of-view, and 
access to longer wavelengths,
and the commissioning and/or launch of powerful observatories
in the X-ray, ultraviolet, infrared, and sub-millimeter, are
but a few of the important technological advances.
This has led to a much-increased empirical understanding of broad
phenomenologies: e.g., constraints on the overall shape of the cosmic history
of star formation, the increased incidence of star-forming
galaxies in less dense environments, the co-evolution 
of stellar bulges and the supermassive black holes that they 
host, and the increasing incidence of galaxy interactions
at progressively higher redshifts, to name but a few.
In turn, tension between these new observational constraints 
and models of galaxy formation and evolution have spurred 
on increasingly complex models, giving important (and sometimes predictive!) 
insight into the physical processes --- star formation (SF),
feedback, galaxy mergers, AGN activity --- that are
driving these phenomenologies.

In this review, I will give a grossly incomplete
and biased overview of some hopefully interesting 
aspects of galaxy assembly.  An important underpinning
of this review is my (perhaps misguided) assumption 
that we live in a Universe whose broad 
properties are described reasonably well by the cold dark 
matter paradigm, with inclusion of a cosmological constant 
($\Lambda$CDM):
$\Omega_{\rm m} = 0.3$, 
$\Omega_{\Lambda} = 0.7$, and $H_0 = 70$\,km\,s$^{-1}$\,Mpc$^{-1}$
following results from the {\it Wilkinson Microwave Anisotropy Probe} 
(\cite[Spergel et al.\ 2003]{spergel03}) and the HST
Key Project distance scale (\cite[Freedman et al.\ 2001]{freedman01}).
This model, while it has important and perplexing fine-tuning problems,
seems to describe detection of `cosmic jerk' using supernova type Ia
(\cite[Riess et al.\ 2004]{riess04}),
and the evolving clustering of the luminous and dark matter content
of the Universe with truly impressive accuracy on a wide range of 
spatial scales (\cite[Seljak et al.\ 2004]{seljak04}).

An important feature of CDM models in general is that galaxies are
formed `bottom-up'; that is, small dark matter halos form first, 
and halo growth continues to the present through 
a combination of essentially smooth accretion and mergers
of dark matter halos (e.g., \cite[Peebles 1980]{peebles80}).  
Thus, small halos were formed very early, whereas larger haloes should still
be growing through mergers.
This has the important feature of decoupling, to a greater or lesser
extent, the physics and timing of the formation of the stars
in galaxies, and the assembly of the galaxies themselves from 
their progenitors through galaxy mergers and accretion 
(\cite[White \& Rees 1978]{white78}; \cite[White \& Frenk 1991]{white91}).

Accordingly, in this review I strive
to explore the two issues separately.  First, I
will explore the build-up of the stellar mass
in the Universe, irrespective of how it is split
up into individual galaxies.  Then, I will move on to
describing some of the first efforts towards understanding 
how the galaxies themselves assembled into their
present forms.  This article will not touch on many 
important and interesting aspects of galaxy evolution;
the co-evolution (or otherwise) of galaxy bulges and 
their supermassive black holes (see, e.g., Peterson, these proceedings;
\cite[Ferrarese et al.\ 2001]{ferrarese01}; \cite[Haehnelt 2004]{haehnelt04}),
the important influences of local environment
on galaxy evolution (see, e.g., \cite[Bower \& Balogh 2004]{bower04}), 
or the evolution of 
galaxy morphology (see, e.g., Franx, these proceedings).  
In this review, I adopt a 
$\Lambda$CDM cosmology and a \cite{kroupa01} IMF; adoption
of a \cite{kennicutt83} IMF in this article would leave the
results unchanged.

\section{The assembly of stellar mass} \label{sec:madau}

In many ways, the empirical exploration of the 
assembly of stellar mass throughout cosmic history
has been one of the defining features of the last decade
of extragalactic astronomical effort.  Yet, in spite of 
such effort, and the apparent simplicity
of the goal, progress at times has been frustratingly
slow.  The difficulties are many:
calibration of SF rate (SFR) and stellar mass 
indicators, the effects of dust on SFR estimates, 
relatively poor sensitivity for most SFR indicators,
and field-to-field variations caused by large scale structure.
In this section, I will briefly summarize 
the progress made to date towards measuring the cosmic 
SF history (SFH) using two independent
methods: exploration of the evolution of the 
cosmic-averaged SFR, and
the evolution of the cosmic-averaged stellar mass density, 
both as a function of epoch.  The two are intimately
related --- the cosmic SFH is the integral over the cosmic SFR ---
and, barring any strong variation in stellar initial mass functions
(IMFs) as a function of galaxy properties and/or epoch, 
consistency between the two would be expected.

\subsection{The cosmic-averaged star formation rate density}

\subsubsection{Methodology}

One measures SFR by measuring galaxy luminosities
in a passband or passbands which one hopes 
will reflect the total number of massive stars in 
a galaxy.  Using stellar population synthesis and
other models, one then attempts to convert this
luminosity into a total SFR, assuming a given 
stellar IMF to allow conversion from the number of
massive stars to the total mass in the newly-formed
stellar population (\cite[see, e.g., Kennicutt 
1998 for an excellent review]{ke98}).

In some cases, this 
calibration from total luminosity to SFR is relatively
robust, because it relies on reasonably well-understood
physics. For example, the total ultraviolet (UV) light from a galaxy
is reasonably robustly translated into the number of massive
O and B stars.  The total infrared luminosity of a starbursting
galaxy, if the galaxy is optically-thick to the UV
light, is a reasonable reflection of the bolometric
output of this starburst.  Balmer line emission, under
weak assumptions about the physical conditions in 
{\sc Hii} regions, is a reasonable indicator of the number
of very massive O stars.  These quantities, in turn, 
can be converted into a total SFR, assuming a stellar IMF
which does not depend on galaxy properties and epoch.

In other cases, the calibration from luminosity to SFR is 
much less direct: e.g., the GHz radio emission from star-forming
galaxies is well-correlated with other indicators of SFR, such 
as IR or Balmer line luminosity, and is known to be dominated
by synchrotron emission from cosmic-ray electrons spiraling
in galactic magnetic fields for at least massive 
star-forming galaxies 
(\cite[see Condon 1992]{condon92} for an excellent review; see also
\cite[Bell 2003]{bell03}).
Yet, there is no robust theoretical understanding 
of why the relationship between radio emission 
and SFR should show such modest scatter (see, e.g.,
\cite[Bressan, Silva, \& Granato 2002; Niklas \& Beck 1997; 
	Lisenfeld, V\"olk, \& Xu 1996]{bressan02,niklas97,lisenfeld96} for
models of the radio emission of star-forming galaxies). 

On top of these interpretive challenges, there are other
important difficulties.  Dust 
extinguishes UV light very effectively; 
empirical dust corrections based
on UV color and/or UV-optical properties calibrated on UV-bright
starbursts in the local Universe (\cite[Calzetti, Kinney, \& 
Storchi-Bergmann 1994; Calzetti 2001]{calzetti94,calzetti01}) 
do not apply to 
normal galaxies (\cite[Bell 2002; Kong et al.\ 2004]{bell02,kong04}), 
IR-bright starbursts 
(\cite[Goldader et al.\ 2002]{goldader02}), 
or indeed even 
{\sc Hii} regions (\cite[Bell et al.\ 2002; Gordon et al.\ 2004]{bellhii,gordon04}).  Estimates of 
dust reddening from Balmer line ratios may yield
reasonably accurate Balmer line-derived SFRs for galaxies
(\cite[Kennicutt 1983]{kennicutt83}), yet are extremely
challenging to measure at $z \gsim 0.4$.  IR and radio 
facilities lack the sensitivity to 
probe to faint limits; only galaxies with SFRs in excess of 
$\sim 10\, {\rm M}_{\odot}\,{\rm yr^{-1}}$ are observable with 
current facilities at 
$z \gsim 0.5$ (e.g., \cite[Flores et al.\ 1999]{flores99}).  

\subsubsection{Results}

There are a huge number of papers which have addressed
the evolution of the cosmic SFR, and are too numerous to 
mention or discuss in any detail.
A few particularly important examples are \cite{lilly96}, 
\cite{madau96}, \cite{flores99}, \cite{steidel99}, and \cite{haarsma00}.

\begin{figure}
  \begin{center}
  \epsfxsize=13.5cm
  \epsfbox{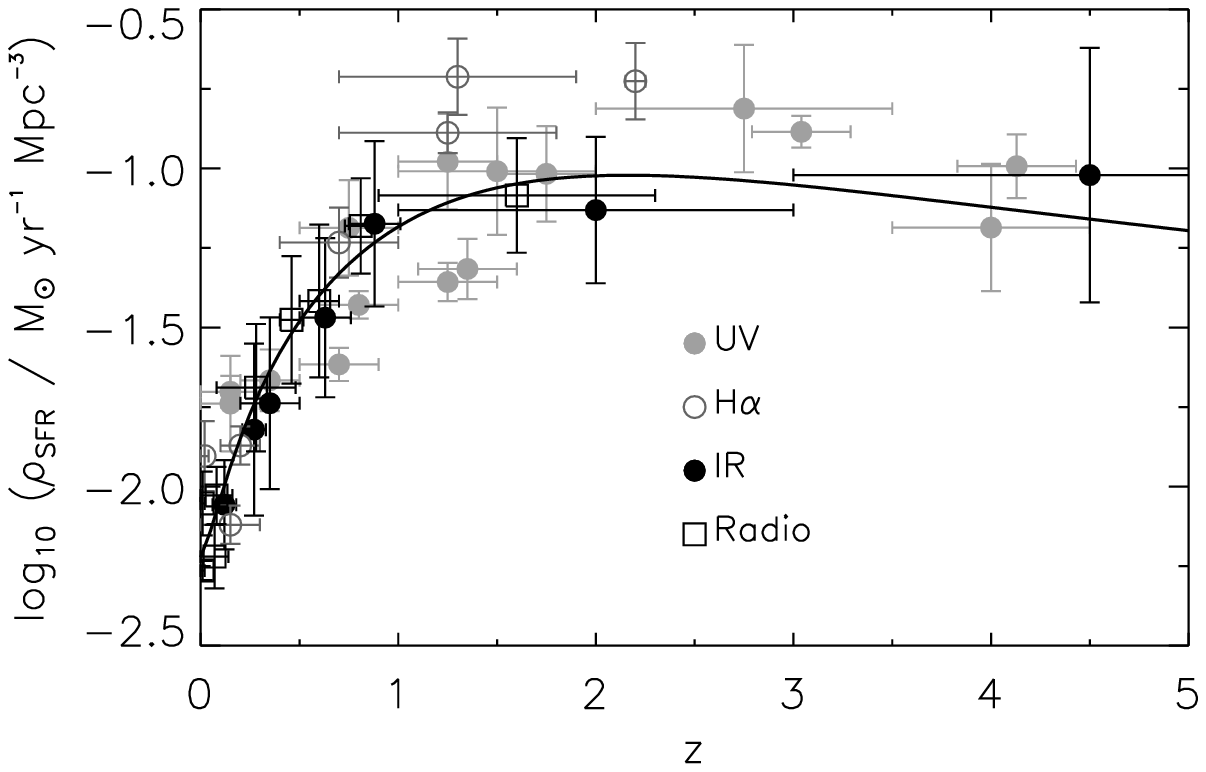}
  \caption{The evolution of the cosmic SFR density.  
  SFRs assume a \protect\cite{kroupa01} IMF and 
  $H_0 = 70$\,km\,s$^{-1}$\,Mpc$^{-1}$.
  The data are taken from \protect\cite{hopkins04}, 
and are corrected for dust assuming a SFR--dust
correlation as found in the local Universe.  The solid
line shows an empirical fit to the cosmic SFR.}\label{fig:sfr}
  \end{center}
\end{figure}

In Fig.\ \ref{fig:sfr}, I show the general form of the 
cosmic SFR, as derived by \cite{hopkins04} in a very
nice compilation of cosmic SFR estimates, where he uses 
locally-calibrated
relationships between dust attenuation and SFR to correct 
for dust.  His corrections are reasonably similar to those commonly used,
but with the important advantages that they are uniformly derived, and 
account for the well-known SFR--dust correlation.
The intention of showing this cosmic SFR is primarily to 
illustrate that despite
the significant observational challenges and interpretive 
challenges it has been possible
to determine its broad shape to better than a factor of three over
much of cosmic history, using a variety of different observational
methodologies.

There are two points which one should take away from the
cosmic SFR, both of which have been known at least at the qualitative level
since 1996 when the first cosmic SFRs were constructed (e.g., 
\cite[Lilly et al.\ 1996; Madau et al.\ 1996]{lilly96,madau96}).
Firstly, it is abundantly clear that the cosmic SFR has dropped by a factor of
nearly 10 since $z \sim 1.5$ (see, e.g., the interesting compilation from 
\cite[Hogg 2001]{hogg01}).  
Secondly, there is no compelling evidence
against a roughly constant cosmic SFR at redshifts higher
than 1 (e.g., \cite[Steidel et al.\ 1999]{steidel99}), 
with the exception of some recent explorations of UV-derived
SFRs at $z \gsim 4$, which appear to be lower
than those at $z \lsim 4$ (e.g., \cite[Stanway et al.\ 2004]{stan04}).

\subsection{The cosmic-averaged star formation history} \label{sec:sfh}

A complementary way to understand the star formation history
of the Universe is to explore the build-up in stellar mass
with cosmic time.  This integral over the cosmic SFR contains
the same information (under the assumption of a reasonably
well-behaved stellar IMF), yet suffers from completely
different systematic uncertainties and is therefore an invaluable
probe of the broad evolution of the stellar content of the Universe.

\subsubsection{Methodology}

Ideally stellar masses would be estimated from spectroscopic
data (e.g., velocity fields, rotation curves, and/or stellar
velocity dispersions) coupled with multi-waveband photometry,
under some set of assumptions about the dark matter content of 
the galaxy.  Unfortunately, measurements of such quality are relatively
uncommon, even in the local Universe.

In the absence of velocity data, one can attempt to estimate stellar 
masses using photometric data alone with the aid of 
stellar population synthesis models (\cite[e.g., Fioc \& Rocca-Volmerange
1997; Bruzual \& Charlot 2003]{pegase,bc03}).  In these models,
increasing the mean stellar age or the metallicity produces
almost indistinguishable effects on their broad-band optical
colors, and indeed even in their spectra with the exception 
of a few key absorption lines (\cite[e.g., Worthey 1994]{worthey94}).
Increasing dust extinction produces very similar
effects at least some of the time (\cite[e.g., Tully et al.\ 1998]{tully98}),
although the relationship between reddening and extinction is 
rather sensitive to star/dust geometry (\cite[Witt \& Gordon 2000]{witt00}).
An increase in stellar population age, metallicity or dust content
reddens and dims the stellar population, at a fixed stellar mass.  Crucially,
the relationship between reddening and dimming is similar 
for all three effects.  While this makes it extremely challenging to measure 
the age, metallicity or dust content of galaxies using 
optical broad-band colors alone, it does mean that one can 
invert the argument and use color and luminosity to rather
robustly estimate stellar mass, almost independent of galaxy SFH,
metallicity or dust content (\cite[e.g., Bell \& de Jong 2001]{ml}).

The slope of the relationship between color and mass-to-light
ratio (M/L) is passband-dependent (steeper in the blue, shallower
in the near-infrared) but does not strongly depend on stellar IMF.  
In contrast, the zero point of the color--M/L relation depends
strongly on stellar IMF, especially to its shape at masses $\lsim 1 M_{\odot}$
where the bulk of the stellar mass resides but the contribution to
the total luminosity is low.
There are important sources of systematic error: 
dust does not always move galaxies along the same
color--M/L relation as defined by age and metallicity
(e.g., \cite[Witt \& Gordon 2000]{witt00}), and most
importantly, significant contributions from young stellar
populations (either because the galaxy is truly young or because
of recent starburst activity) can bias the stellar M/Ls
at a given color towards lower values.  Recently, a number of 
methodologies have been developed to address these 
limitations by inclusion of important variations in
star formation history (\cite[e.g., Papovich, Dickinson, \& Ferguson 
2001]{papovich01})
or by using spectral indices and colors to account for 
bursts and dust more explicitly 
(\cite[e.g., Kauffmann et al.\ 2003]{kauffmann03}).

\subsubsection{Results}

\begin{figure}
  \begin{center}
  \epsfxsize=13.5cm
  \epsfbox{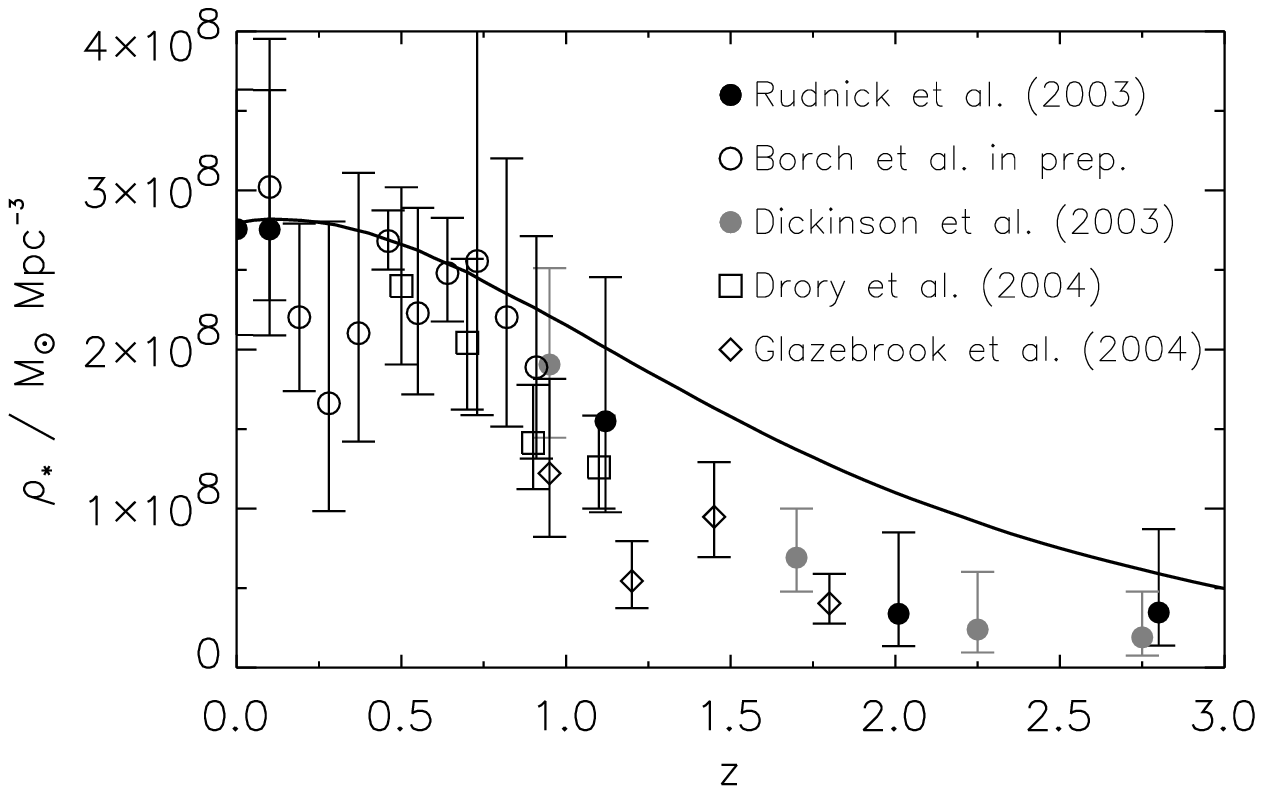}
  \caption{The evolution of the cosmic-averaged stellar mass density.
  Stellar masses assume a \protect\cite{kroupa01} IMF and 
  $H_0 = 70$\,km\,s$^{-1}$\,Mpc$^{-1}$.
  The data are taken from \protect\cite[Rudnick et al.\ (2003; normalized
to reproduce the $z = 0$ stellar mass density from Cole et al.\ 2001 \& Bell 
et al.\ 2003)]{rudnick03,cole01,bellmf},
  \protect\cite{dickinson03}, \protect\cite{drory04}, 
  \protect\cite{glaze04}, and 
  Borch et al.'s (in preparation)
  measurements from the COMBO-17 photometric redshift survey.
  The solid
line shows the integral of the SFR from Fig.\ \ref{fig:sfr}.}\label{fig:sfh}
  \end{center}
\end{figure}

The basic methodology has been applied in the last several years to a 
wide variety of galaxy surveys to explore the stellar mass density in 
the local Universe (e.g., \cite[Cole et al.\ 2001]{cole01}; 
\cite[Bell et al.\ 2003]{bellmf}), its evolution out to $z \sim 1$ 
(e.g., \cite[Brinchmann \& Ellis 2000]{brinchmann00}; 
\cite[Drory et al.\ 2004]{drory04};
Borch et al., in preparation) and even out to $z \sim 3$
(e.g., \cite[Rudnick et al.\ 2003]{rudnick03}; 
\cite[Dickinson et al.\ 2003]{dickinson03}; 
\cite[Fontana et al.\ 2003]{fontana03}, 
\cite[Glazebrook et al.\ 2004]{glaze04}).
In Fig.\ \ref{fig:sfh}, I show a compilation of 
some of these stellar mass estimates, 
all transformed to a \cite{kroupa01} IMF.  The error bars in 
all cases give a rough idea of the uncertainties in 
stellar mass; the error bars for \cite{drory04}, \cite{glaze04}, and 
Borch et al.\ include contributions from 
cosmic variance also.

From Figs.\ \ref{fig:sfr} and \ref{fig:sfh}, it is clear
that the epoch $z \gsim 1$ is characterized by rapid
star formation, with roughly 2/3 of the stellar mass
in the Universe being formed in the first 5 Gyr.  In contrast,
from $z \sim 1$ to the present day, one witnesses a striking 
decline in the cosmic SFR, by a factor of roughly 10.  
Despite this strongly suppressed SFR, roughly 1/3 of all
stellar mass is formed in this interval, owing to the large
amount of time between $z \sim 1$ and the present day.  

\subsection{Are the cosmic star formation rates and histories consistent?}

It is interesting to test for consistency between the 
cosmic SFR and cosmic SFH --- a lack of consistency between
these two may give important insight into the form of the 
stellar IMF between $\sim 1 M_{\odot}$ (the mass range which
optical/NIR light is most sensitive to) and $\gsim 5 M_{\odot}$
(the stellar mass range probed by SFR indicators), or 
the calibrations of or systematic errors in stellar mass and/or SFR
determinations. 

In order to integrate the cosmic SFR, I choose to roughly
parameterize the form of the cosmic SFR following 
\cite{cole01}.  The cosmic SFR $\psi = 
(0.006+0.072 z^{1.35})/[1+(z/2)^{2.4}] M_{\odot} {\rm yr^{-1}\,Mpc^{-3}}$
provides a reasonable fit to the cosmic SFR (Fig.\ \ref{fig:sfr}).  
The data are insufficient
to constrain whether or not the cosmic SFR declines towards high redshifts
from a maximum at $z \sim 1.5$; I choose to impose a mild decrease
towards high redshift, primarily because that matches the 
evolution in integrated stellar mass slightly better than a 
flat evolution.  This cosmic SFR is then integrated using the 
PEGASE stellar population model assuming a \cite{kroupa01} IMF
and an initial formation redshift $z_f = 5$.  
The exact integration in PEGASE accounts explicitly for
the recycling of some of the initial stellar bass back into 
the interstellar medium; for the \cite{kroupa01} 
IMF this fraction is $\sim 1/2$, i.e., stellar mass in long-lived
stars is 1/2 of the stellar mass initially formed
(for a \cite[Kennicutt 1983]{kennicutt83} IMF the fraction is 
similar).
I show the expected cosmic SFH as the solid line in 
Fig.\ \ref{fig:sfh}.  

It is clear that the form of the cosmic SFR required in
Fig.\ \ref{fig:sfr} reproduces rather well the cosmic SFH
as presented in Fig.\ \ref{fig:sfh}.  There are some slight
discrepancies; a cosmic SFR as flat as that shown in Fig.\ \ref{fig:sfr}
appears to overpredict the amount of stellar mass
that one sees at $z \sim 3$.  This might indicate 
that a drop-off in cosmic SFR towards higher
redshift is required, or may indicate that an IMF richer in 
high-mass stars is favored for high-redshift starbursts.
Yet, it is important to remember that estimates
of cosmic SFR and SFH are almost
impossible to nail down with better than 30\% accuracy at any redshift.
At $z \gsim 1$ the constraints are substantially weaker still, 
owing to large uncertainties from large scale structure, uncertainties
in the faint-end slope of the stellar mass or SFR functions used
to extrapolate to total SFRs or masses, and the difficulty
in measuring SFRs and masses of intensely star-forming, dusty
galaxies.   Therefore, I would tend to downweight this disagreement
at $z \gsim 1.5$ until better and substantially deeper 
data are available, focusing instead on the rather pleasing
overall agreement between these two independent probes of the 
cosmic SFH.

\section{The assembly of galaxies}

\S \ref{sec:madau} focused on the broadest possible picture 
of the assembly of stellar mass --- the build-up of
the stellar population, averaged over cosmologically-significant
volumes.  Yet, the physical processes contributing to this evolution
will be much more strongly probed by studying the demographics of 
the galaxy populations as they evolve.  This splitting of the 
cosmic SFR/SFH can happen in many ways: study of the evolution 
of the luminosity function, stellar mass function, or SFR `function',
study of galaxies split by morphological type or rest-frame
color, or by identification of galaxies during particular phases
of their evolution (e.g., interactions).  There has been a great 
deal of activity over the last decade towards this goal:
implicit in the exploration of Figs.\ \ref{fig:sfr} and \ref{fig:sfh}
is the construction of SFR and stellar mass functions,
and a number of studies have explored the evolution 
of the galaxy population split by morphological type 
(\cite[e.g., Brinchmann \& Ellis 2000]{brinchmann00}, 
\cite[Im et al.\ 2002]{im02}), rest-frame color 
(\cite[e.g., Lilly et al.\ 1996]{lilly96},
\cite[Wolf et al.\ (2003)]{wolf03},
\cite[Bell et al.\ 2004b]{bell04}), or focusing on the role
of galaxy interactions (\cite[e.g., Le F\`evre et al.\ 2000]{lefevre00},
\cite[Patton et al.\ 2002]{patton02}, 
\cite[Conselice et al.\ 2003]{conselice03}).  

A full and fair exploration of any or all of these goals
is unfortunately beyond the scope of this work.  Here, I choose
to focus on one particular key issue: the importance of galaxy mergers
in driving galaxy evolution in the epoch since $z \sim 3$, and 
especially at $z \lsim 1$.  Unlike the evolution of e.g., the stellar
mass function or SFR function, to which
both quiescent evolution and galaxy accretion can contribute, 
galaxy mergers (especially those at $z \lsim 2$) are an unmistakable
hallmark of the hierarchical assembly of galaxies.  Therefore,
exploration of galaxy mergers directly probes one of the key
features of our current cosmological model.   

I will focus here on exploring the number of major galaxy 
mergers (traditionally defined as those with mass ratios 
of 3:1 or less) over the last 10 Gyr.
There are three complementary approaches to exploring 
merger rate, all of which suffer from important
systematic uncertainties: the evolution of the fraction 
of galaxies in close pairs, the evolving fraction of 
galaxies with gross morphological irregularities, and 
investigation of the evolution of plausible merger
remnants.  In this work, I will briefly discuss
all three methods, highlighting areas of particular 
uncertainty to encourage future development in this 
exciting and important field.

\subsection{The evolution of close galaxy pairs} \label{sec:close}

\begin{figure}
  \begin{center}
  \epsfxsize=12cm
  \epsfbox{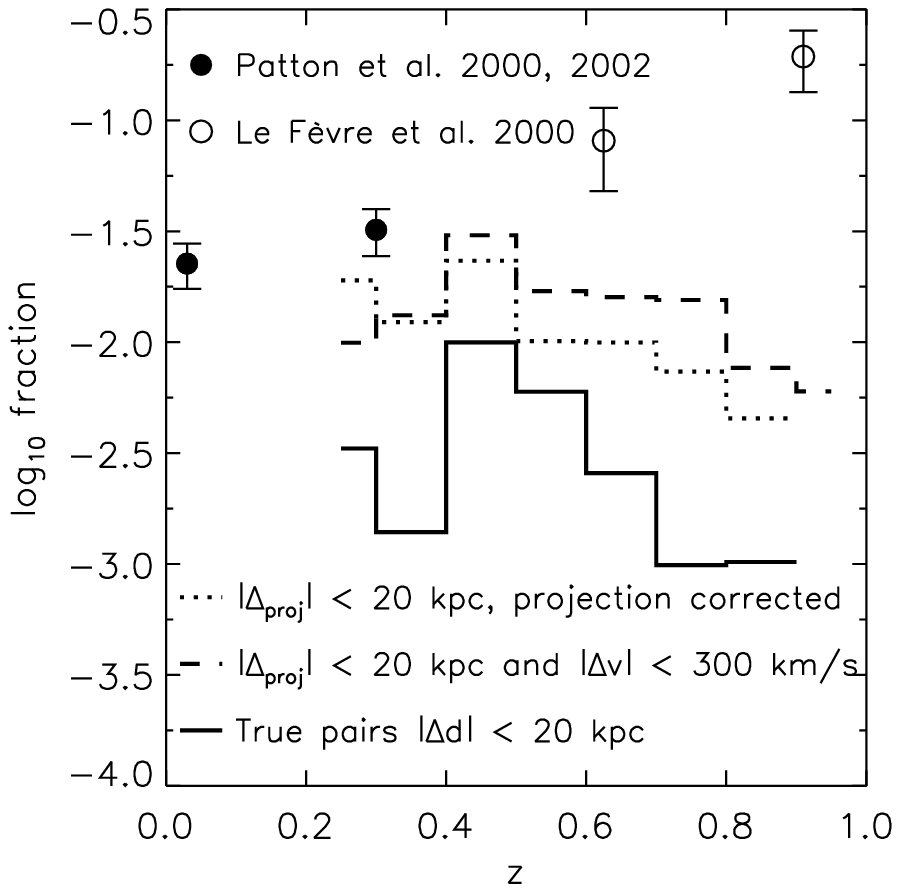}
  \caption{The evolution of the close pair fraction.
	Data points show measurements of the fraction of 
	galaxies with 1 or more neighbors within a projected
	distance of 20$h^{-1}$\,kpc, corrected for projection (open
	points; \cite[Le F\`evre et al.\ 2000]{lefevre00}), and
	within a projected distance of 20$h^{-1}$\,kpc and a velocity
	separation of $< 500$\,kms$^{-1}$ (solid
	points; \cite[Patton et al.\ 2000, 2002]{patton00,patton02}).
	The lines show the evolution of approximately 
	equivalent measurements from the van Kampen et al.\ (in
	prep.) mock COMBO-17 catalogs: projection-corrected
	fraction (dotted line), the fraction of 
	galaxies with $\ge 1$ neighbor within 20$h^{-1}$\,kpc
	and with velocity difference $< 300$\,kms$^{-1}$ (dashed
	line), and the real fraction of galaxies with $\ge 1$
	neighbor within 20$h^{-1}$\,kpc in real space.
}\label{fig:close}
  \end{center}
\end{figure}

One of the most promising measures of galaxy merger rate 
is the evolution in the population of close galaxy pairs.
Most close (often defined as being separated by $< 20$\,kpc),
bound galaxy pairs with roughly equal masses will 
merge within $\lsim 1$ Gyr owing
to strong dynamical friction (e.g.,  \cite[Patton et al.\ 2000]{patton00}). 
Thus, if it can be measured, the fraction of galaxies
in physical close pairs is an excellent proxy for 
merger rate.  

Accordingly, there have been a large number of studies
which have attempted to measure close pair fraction evolution
(a few examples are \cite[Zepf \& Koo 1989]{zepf89}, 
\cite[Carlberg, Pritchet, \&
Infante 1994]{carlberg94}, \cite[Le F\`evre et al.\ 2000]{lefevre00}; see 
\cite[Patton et al.\ 2000]{patton00} for an extensive discussion
of the background to this subject).  In Fig.\ \ref{fig:close}, 
I show the fraction of $M_B - 5\log_{10}h \lsim -19.5$ galaxies in close 
$|\Delta_{\rm proj}| < 20 h^{-1}$\,kpc pairs from \cite{patton00}, 
\cite{lefevre00}, and \cite{patton02}.
\cite{patton00} and \cite{patton02} explore the fraction of galaxies
in close pairs with velocity differences $< 500$\,kms$^{-1}$, 
whereas \cite{lefevre00} study projected galaxy pairs, 
and correct them for projection in a statistical way.
Both studies, and indeed most other studies, 
paint a \textit{broadly} consistent picture that the frequency of galaxy 
interactions was much higher in the past than at the present, 
and that the average $\sim L^*$ galaxy has suffered
0.1--1 major interaction between $z \sim 1$ and the present day.
Small number statistics, coupled with differences in assumptions
about how to transform pair fractions into merger rate, 
lead to a wide dispersion in the importance of major merging 
since $z \sim 1$.

Yet, there are significant obstacles to the interpretation 
of these, and indeed future, insights into 
galaxy major merger rate evolution.  Technical
issues, such as contamination from foreground or background galaxies, 
redshift incompleteness, and the construction of equivalent samples
across the whole redshift range of interest must be thought 
about carefully (see, e.g., 
\cite[Patton et al.\ 2000, 2002]{patton00,patton02}). 
An further concern is that most close pair 
measurements are based on galaxies with similar 
magnitudes in the rest-frame optical.  Bursts of 
tidally-triggered star formation may enhance the optical
brightness of both galaxies in pairs, and the selection
of galaxies with nearly equal optical brightness may not 
select a sample of galaxies with nearly equal stellar masses.
\cite{bundy04} used $K$-band images of a sample of 190
galaxies with redshifts to explore this source of bias, 
finding a substantially lower fraction of galaxies in 
nearly equal-mass close pairs than derived from 
optical data.  

Yet, it is also vitally important to build one's intuition
as to how the measured quantities relate to the true 
quantities of interest.  
As a crude example of this, I carry
out the simple thought experiment where we
compare the pair fraction derived from projection-corrected
projected pair statistics (dotted line in Fig.\ \ref{fig:close}), 
the pair fraction derived
if one used spectroscopy to keep only those galaxies with 
$|\Delta v| < 300$\,kms$^{-1}$ (dashed line), and the true fraction of 
galaxies with real space physical separation of $< 20h^{-1}$\,kpc
(solid line).  I use a mock 
COMBO-17 catalog under development by van Kampen et al.\ (in prep.; see
\cite[Van Kampen, Jimenez, \& Peacock 1999]{vankampen99} for
a description of this semi-numerical technique)
to explore this issue; this model is used with their kind
permission.  It is important to note
that this simulation is still under development; 
the match between the observations and models
at $z \lsim 0.5$ is mildly encouraging, although the 
generally flat evolution of pair fraction may not be 
a robust prediction of this model.  Nonetheless, this
model can be used to great effect to gain some insight into sources of bias
and uncertainty.

The catalog is tailored to match 
the broad characteristics of the COMBO-17 survey to the largest
extent possible (see \cite[Wolf et al.\ 2003]{wolf03} for a description
of COMBO-17): it has $3 \times 1/4$ square degree fields, limited
to $m_R < 24$ and with photometric redshift accuracy mimicking COMBO-17's 
as closely as possible.  Primary galaxies with $m_R < 23$, 
$0.2 \le z \le 1.0$ and $M_V \le -19$ are chosen; satellite 
galaxies are constrained only to have $m_R < 24$ and 
$|\Delta m_R| < 1$\,mag (i.e., to have a small luminosity
difference).  The dotted and dashed lines
show the pair fraction recovered by approximately reproducing
the methodologies of \cite[Le F\`evre et al.\ (2000; dotted line)]{lefevre00}
and \cite[Patton et al.\ (2000, 2002; dashed line)]{patton00,patton02}.
It is clear that statistical field subtraction, adopted
by \cite{lefevre00} and others, may be rather robust, in terms of 
recovering the trends that one sees with the spectroscopic$+$imaging 
data.  The solid line shows the fraction of galaxies 
actually separated by 20$h^{-1}$\,kpc.  It is clear that, 
at least in this simulation,
many galaxies with projected close separation and small 
velocity difference are members of the primary galaxy's group which
happen to lie close to the line of sight to the primary galaxy but
are $\gsim 20$ kpc from the primary.  In \cite{lefevre00}, \cite{patton02}, 
and other studies, this was often corrected for by multiplying
the observed fraction by $0.5(1+z)$ following the analysis of \cite{yee95};
the data showed in Fig.\ \ref{fig:close} was corrected in this way.
Our analysis of these preliminary COMBO-17 mock catalogs 
suggest that these corrections are uncertain, 
and that our current understanding 
of galaxy merger rate from close pairs may be biased.
It is important to remember that the simulations discussed
here are preliminary; the relationship between `observed' and 
true close pair fractions may well be different from the trends
predicted by the model.  Yet, it is nonetheless clear
that further modeling work is required before one can 
state with confidence that one understands the implications 
of close pair measurements.

\subsection{The evolution of morphologically-disturbed galaxies}
\label{sec:mergers}

With the advent of relatively wide-area, high spatial
resolution and S/N imaging from HST, it has become feasible
to search for galaxy interactions by identifying morphologically-disturbed
galaxies.  There are a number of ways in which 
one can evaluate morphological disturbance: a few 
examples are by visual 
inspection (e.g., \cite[Arp 1966]{arp66}), 
residual structures in unsharp-masked or model-subtracted
images (e.g., \cite[Schweizer \& Seitzer 1992]{schweizer92}),
automated measures of galaxy asymmetry
(e.g., \cite[Conselice, Bershady, \& Jangren 2000]{con00}), or by 
the distribution of pixel brightnesses and 2nd-order moment
of the light distribution 
(e.g., \cite[Lotz, Primack, \& Madau 2004]{lotz04}).

The important hallmarks that these
different methodologies probe are the non-equilibrium 
signatures of tidal interaction --- 
multiple nuclei and extended tidal tails.
These features are common in simulations
of interacting galaxies, and cannot result from
quiescent secular evolution (e.g., \cite[Toomre \& Toomre 1972]{toomre72},
\cite[Barnes \& Hernquist 1996]{barnes96}).
Visual inspection is a subjective way to pick out these
structures, even to relatively low surface brightness limits; 
in contrast, automated measures of morphology
typically quantify the amount of light in bright asymmetric
structures, so pick out multiple 
bright patches or bright tidal tails rather effectively.

\begin{figure}
  \begin{center}
  \epsfxsize=12cm
  \epsfbox{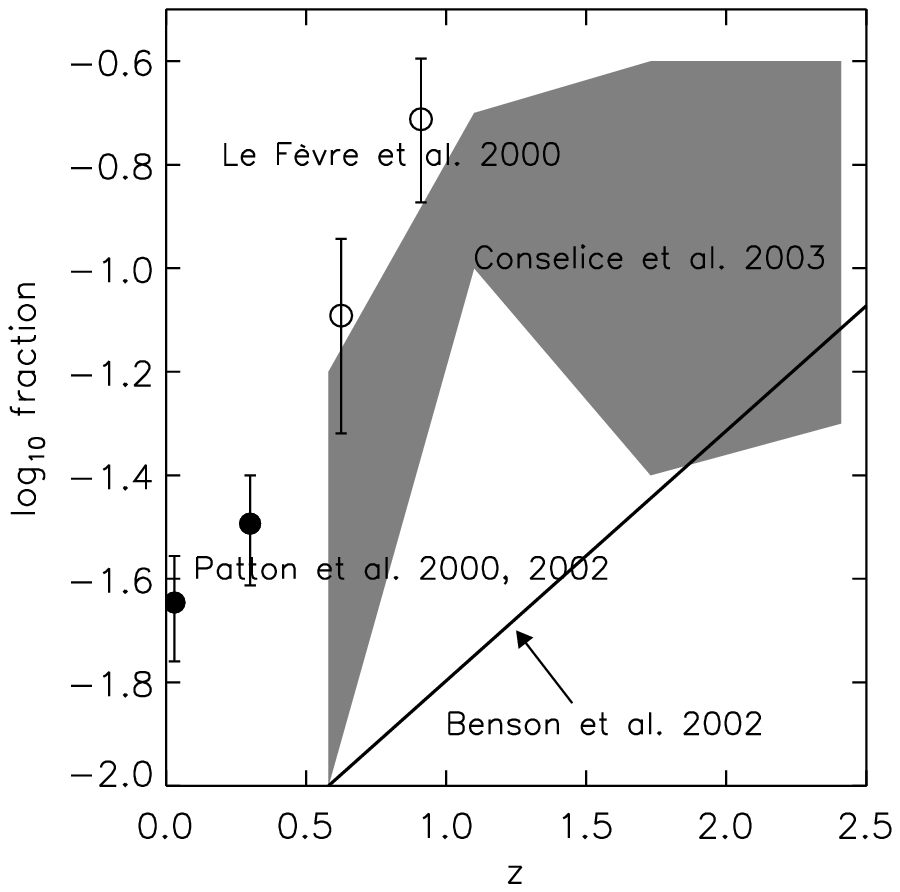}
  \caption{The evolution of merger rate as inferred
	from the fraction of galaxies with gross asymmetries
	(\cite[Conselice et al.\ 2003]{conselice03}; grey
	shaded region).  Inferred merger rates from 
	close pairs are reproduced from Fig.\ \ref{fig:close}
	for comparison.  A rough fit to predictions of major merger rate
	(visible for 1 Gyr) from the 
	semi-analytic galaxy formation model of
	\cite{benson02}, reproduced from \cite{conselice03}, 
	is shown by the solid line for reference.
}\label{fig:asymm}
  \end{center}
\end{figure}

Recently, \cite{conselice03} have explored the fraction 
of galaxies with strong asymmetries in the rest-frame $B$-band
our to redshift $\sim 3$ in the Hubble Deep Field North.
Their results, for a similar absolute magnitude
range as explored by \cite{patton02} and \cite{lefevre00},
are shown in Fig.\ \ref{fig:asymm}.  While small number
statistics (there are less than 500 galaxies contributing to the
fractions), coupled with asymmetry uncertainties, are clearly
an issue, a broadly consistent picture is painted whereby 
galaxy mergers were substantially more frequent at $z \gsim 1$, 
and have decreased in frequency until the present day.  It is
interesting to note that the observations are not well-reproduced
by the Benson et al.\ or van Kampen et al.\ models, which both 
predict substantially lower merger rates.  Indeed, perhaps for
the first time, observers are invoking the need for many more
mergers than theorists!

While these first tentative steps are encouraging, there are
a number of systematic uncertainties that should be considered
carefully.  At some level, contamination from
projected close galaxy pairs is bound to be a problem.  Samples
of highly asymmetric galaxies will, correctly, include also some
very close physically-associated galaxy pairs.  Unfortunately, they
will also contain a number of physically-unassociated close 
galaxy pairs: \cite{lefevre00} find that only $\sim 30$\% of 
close galaxy pairs are likely to be physically-associated
(in the sense of being in the same galaxy group; a smaller
fraction still are genuine close pairs), and the mock COMBO-17 catalog
analysis presented earlier suggests a fraction closer to 10\%.  
These contaminants may be weeded out by source extraction 
software, as the pair of galaxy images may be parsed into 
2 separate galaxies, each of which has a low asymmetry.  Yet, 
if this image parsing is too enthusiastic, genuine major interactions
may be parsed into multiple, individually rather symmetric, sub-units.

Furthermore, there are differences between different
classification methodologies as to what constitutes
a `merger'.  An example of this is given in Fig.\ 18 of 
\cite{conselice04}, who explore the distribution of 
morphologically early-type galaxies (those with
dominant bulge components), late-type galaxies (those 
with dominant disks), and peculiar galaxies (all other
types, which includes merging and irregular galaxies)
in the concentration (C) and asymmetry (A) plane.  
C is a measure of the light concentration of a galaxy profile, 
having high values for centrally-concentrated light profiles, and
A is a measure of asymmetry, with high values denoting asymmetric
galaxies.  Inspecting their Fig.\ 18, one sees
scattered trends between visual and automated
classifications.  Early-type galaxies tend to have higher concentrations
and lower asymmetries, later types have lower concentrations and 
lower asymmetries, and peculiar systems have low concentrations
and a wide range of asymmetries, with a tail to very high asymmetries.
Yet, there are a number of early-types with low concentrations, 
and importantly a large number of peculiar galaxies have low asymmetries,
and cannot be distinguished from morphologically-normal galaxies in 
this scheme.  Preliminary comparisons between by-eye morphologies and 
C and A values for F606W images of galaxies in 
the GEMS (Galaxy Evolution from Morphology and SEDs; 
\cite[Rix et al.\ 2004]{rix04}) HST survey supports this
conclusion, indicating
that such automated schemes find it hard to differentiate between
irregular galaxies (whose morphologies, to the human eye, 
indicate sporadic star formation triggered by internal processes)
and clearly interacting galaxies with morphological features
indicative of tidal interactions, such as multiple bright 
nuclei or tidal tails.  A further complication is that the human
eye is more sensitive to faint tidal tails in early- or late-stage mergers
than automated schemes (i.e.,
the timescale probed by visual classifications may be longer than
for automated schemes).

This discussion is by no means meant to detract from the 
value of the intriguing and ground-breaking work 
on merger rates from 
morphology; rather it is to emphasize that the measurement
of merger rate is a subtle and difficult endeavor, fraught with 
systematic uncertainties which will likely have to be modeled
explicitly using future generations of $N$-body, semi-analytic and
SPH simulations, coupled with the artificial redshifting experiments
already commonly used in this type of work.

\subsection{The evolution of plausible merger remnants}  \label{sec:early}

From the earliest days of galaxy morphological classification,
a population of galaxies whose light distribution is dominated
by a smoothly distributed, spheroidal, centrally-concentrated  
light distribution was noticed.  These \textit{early-type 
galaxies} are largely supported by the random motions of their
stars (\cite[Davies et al.\ 1983]{davies83}).  These properties
are very naturally interpreted as being the result of violent
relaxation in a rapidly-changing potential well (\cite[Eggen, 
Lynden-Bell, \& Sandage 1962]{eggen62}).  Therefore, in our present
cosmological context, these galaxies are readily identified with the 
remnants of major galaxy mergers (e.g., 
\cite[Toomre \& Toomre 1972]{toomre72}; 
\cite[Barnes \& Hernquist 1996]{barnes96}).  Detailed comparisons
of simulated major merger remnants broadly supports
this notion although some interesting discrepancies 
with observations remain, and are perhaps
telling us about difficult-to-model gas-dynamical 
dissipative processes (e.g., \cite[Bendo \& Barnes 2000]{bendo00};
\cite[Meza et al.\ 2003]{meza03}; \cite[Naab \& Brukert 2003]{naab03}).
There is strong observational support for this notion ---  
the correlation between fine morphological structure and 
residuals from the color--magnitude correlation 
(\cite[Schweizer \& Seitzer 1992]{schweizer92}), the existence of 
kinematically-decoupled cores (e.g., \cite[Bender 1988]{bender88}), and 
the similarity between the stellar dynamical properties
of late-stage IR-luminous galaxy mergers and elliptical galaxies
(e.g., \cite[Genzel et al.\ 2001]{genzel01}).

Therefore, study of the evolution of spheroid-dominated, early-type
galaxies may be able to give important insight into the importance
of galaxy merging through cosmic history.  There are, as always, 
a variety of 
important complications and limitations to this approach.  For example, 
a fraction of the galaxies 
becoming early-type during galaxy mergers will later re-accrete a 
gas disk, which will gradually transform into stars, making the 
galaxy into a later-type galaxy with a substantial bulge component
(e.g., \cite[Baugh, Cole, \& Frenk 1996]{baugh96}).  Furthermore, 
not all galaxy mergers
will result in a spheroid-dominated galaxy; some lower mass-ratio
interactions will result in a disk-dominated galaxy (e.g., 
\cite[Naab \& Burkert 2003]{naab03}).  In addition, it would 
be foolish to \textit{a priori} ignore the possibility that 
an important fraction of early-type galaxies may be formed rapidly
in mergers of very gas-rich progenitors at early epochs ---
a scenario reminiscent of the classical monolithic
collapse picture (\cite[Larson 1974]{larson74};  
\cite[Arimoto \& Yoshii 1987]{arimoto87}).
Yet, these interpretive complications, much as in all the cases
discussed above, do not lessen the value of placing 
observational constraints on the phenomenology, with the confidence that
our interpretation and understanding of the phenomenology will
improve with time.

The rate of progress in this field has largely been determined
by the availability of wide-format high-resolution imaging from HST. 
Ground-based resolution is insufficient to robustly distinguish 
disk-dominated and spheroid-dominated galaxies at cosmologically-interesting
redshifts.  In the local Universe, the vast majority of 
morphologically early-type galaxies occupy a relatively tight
locus in color--magnitude space --- the color--magnitude relation 
(e.g., \cite[Sandage \& Visvanathan 1978]{sandage78};
\cite[Bower, Lucey, \& Ellis 1992]{ble}; 
\cite[Schweizer \& Seitzer 1992]{schweizer92}; 
\cite[Strateva et al.\ 2001]{strateva01}).  Therefore, 
many workers have focused on the evolution of the 
red galaxy population as an accessible alternative.
The efficacy of this approach is only recently being
tested.  Accordingly, I explore the evolution of the 
red galaxy population first, turning subsequently 
to the evolution of the early-type population later.

\subsubsection{The evolution of the total stellar 
mass in red-sequence galaxies}

It has become apparent only in the last 5 years that the color
distribution of galaxies is bimodal in both the local 
Universe (e.g., \cite[Strateva et al.\ 2001]{strateva01}; \cite[Baldry et al.\
2004]{baldry04}) and out to $z \sim 1$ (\cite[Bell et al.\ 2004b]{bell04}).
This permits a model-independent definition of red galaxies ---
those that lie on the color--magnitude relation.  A slight
complication is that the color of the red sequence evolves with 
time as the stars in red-sequence galaxies age, necessitating 
an evolving cut between the red sequence and the `blue cloud'.  
The evolution of this cut means that 
some workers who explore the evolution of red
galaxies using rather stringent color criteria --- e.g., 
galaxies the color of local E/S0 galaxies --- find much faster
evolution in the red galaxy population than those who adopt 
less stringent or evolving cuts (e.g., \cite[Wolf et al.\ 2003]{wolf03}).

\begin{figure}
  \begin{center}
  \epsfxsize=13.5cm
  \epsfbox{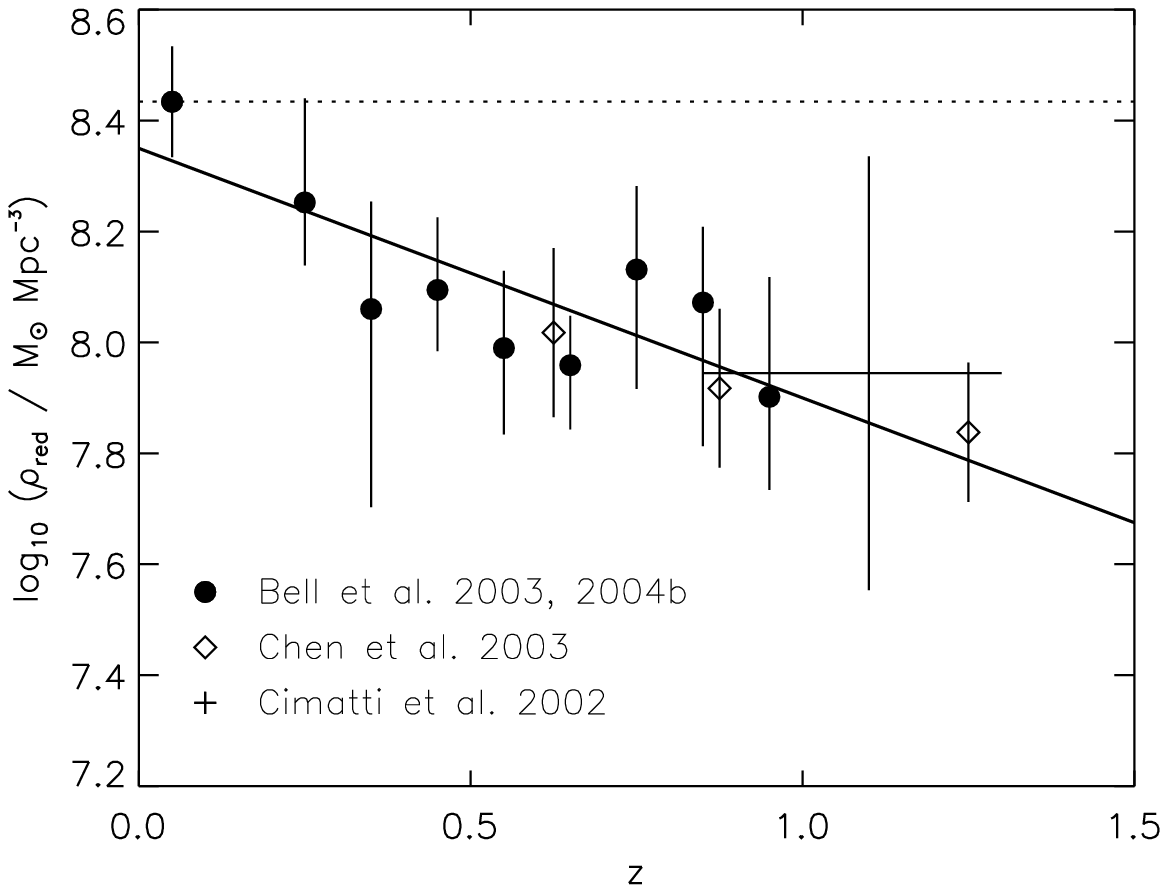}
  \caption{The evolution of the stellar mass density in 
  red-sequence galaxies.
  Stellar masses assume a \protect\cite{kroupa01} IMF and 
  $H_0 = 70$\,km\,s$^{-1}$\,Mpc$^{-1}$.
  The local data point is taken from \protect\cite{bell03},
  and data for $0.2 < z < 1.3$ is taken from \protect\cite{bell04},
  \protect\cite{chen03}, and \protect\cite{cimatti02}.
  The solid
line shows a rough fit to the total stellar mass density in 
red-sequence galaxies in the \protect\cite{cole00} semi-analytic
galaxy formation 
model.  The dotted line shows the expected result
if red-sequence galaxies were formed at $z \gg 1.5$ and simply
aged to the present day.  }\label{fig:red}
  \end{center}
\end{figure}

Bearing this in mind, I show the evolution in the total stellar mass
in red galaxies in Fig.\ \ref{fig:red}.  Although many surveys could
in principle address this question (e.g., the surveys discussed
in \S \ref{sec:sfh}), very few 
have split their stellar mass evolution by color, 
and to date most surveys have only published evolution of luminosity densities
in red galaxies.  The $z = 0$ stellar mass density in red
galaxies is from SDSS and 2MASS (\cite[Bell et al.\ 2003]{bellmf}).
The COMBO-17 datapoints for $j_B$ evolution 
(\cite[Bell et al.\ 2004b]{bell04}) are converted to stellar mass
using color-dependent stellar M/Ls as defined by \cite{ml};
extrapolation to total stellar mass density adopts a
faint-end slope $\alpha = -0.6$, as found by 
\cite[Bell et al.\ (2003; 2004b)]{bellmf,bell04} for red-sequence
galaxies at $z \lsim 1$.  
Error bars account for stellar mass uncertainties and cosmic variance, 
defined by the field-to-field scatter in stellar mass densities 
from the 3 COMBO-17 fields.
Stellar masses for the LCIRS sample of red galaxies
were estimated using the rest-frame $R$-band 
luminosity density presented by \cite{chen03}, accounting for 
the mildest possible passive luminosity evolution (to account 
for evolution in stellar population 
color and luminosity in the most conservative way, so as to minimize
any stellar mass evolution), and using a stellar M/L for early-type galaxies
from \cite{ml} using a \cite{kroupa01} IMF, and adopting
a color of $B-R = 1.5$ for early-type galaxies as a $z = 0$ baseline.
Error bars for include stellar 
mass estimation uncertainties and estimated cosmic
variance, following \cite{somerville04}.
The K20 data point at $z \sim 1.1$, derived from ERO `old' galaxy
space densities from \cite{cimatti02}, is very roughly calculated
using a number of doubtless poor assumptions. A (rather large) stellar mass
of $\sim 10^{11} M_{\odot}$ is attached to each galaxy, 
and the densities are multiplied by 2 to account for the star-forming
EROs (the split was roughly 50:50).  This stellar mass density
was multiplied by 2 again to account for fainter, undetected
galaxies.  Error bars of $\pm$0.2 dex and $\pm$0.3 dex, combined
in quadrature, account for cosmic variance following \cite{somerville04} plus our poor modeling 
assumptions.   Owing to their use of discordant cosmologies, I do not 
show the inferred stellar mass evolution of the CFRS
red galaxies in Fig.\ \ref{fig:red}; however, like
\cite{bell04} they infer no evolution in the rest-frame
$B$-band luminosity density to within their errors 
(\cite[Lilly et al.\ 1995]{lilly95}), meaning
that their stellar mass evolution would fall into excellent
agreement with those of \cite{bell04} or \cite{chen03}.

The results are shown in Fig.\ \ref{fig:red}.  To first order, the 
luminosity density in the optical in red galaxies is constant out
to $z \sim 1$; this is confirmed by a number of surveys (e.g., 
\cite[Lilly et al.\ 1995]{lilly95}, \cite[Chen et al.\ 2003]{chen03},
or \cite[Bell et al.\ 2004b]{bell04}).  Coupled with the passive
ageing of the stellar populations of these red galaxies (as 
is indicated by their steady reddening with cosmic time, and 
is confirmed by study of dynamically-derived M/Ls
and absorption line ratios; e.g., \cite[Wuyts et al.\ 2004]{wuyts04};
\cite[Kelson et al.\ 2001]{kelson01}), this 
implies a steadily increasing stellar mass density in 
red galaxies since $z \sim 1.2$.  \textit{To date, to the 
best of our knowledge, there are no published determinations
of red galaxy stellar mass or luminosity 
density which contradict this picture}.

The implications of this result are rather far-reaching.  
Bearing in mind that at $z \gsim 1$ that the red galaxy population may 
be significantly contaminated and/or dominated by dusty
star-forming galaxies (e.g., \cite[Yan \& Thompson 2003]{yan03},
\cite[Moustakas et al.\ 2004]{moustakas04}), this evolution may well
represent a strong upper limit to the stellar mass in 
early-type galaxies since $z \sim 1.2$, unless large populations
of very bright blue morphologically early-type galaxies 
are found.  I address this question next, by exploring
the stellar mass evolution in morphologically early-type
galaxies since $z \sim 1$.

\subsubsection{The evolution of the total stellar mass in early-type galaxies}

Owing to the lack of extensive wide-area HST-resolution imaging data,
there are even weaker constraints on the evolution of stellar
mass in morphologically early-type galaxies.
We show published results from \cite{im02}, \cite{conselice04}, 
and \cite{cross04}, supplementing them with 
a preliminary analysis of galaxies from 
GEMS and COMBO-17, which is presented with the permission 
of the GEMS and COMBO-17 teams. 

\begin{figure}
  \begin{center}
  \epsfxsize=13.5cm
  \epsfbox{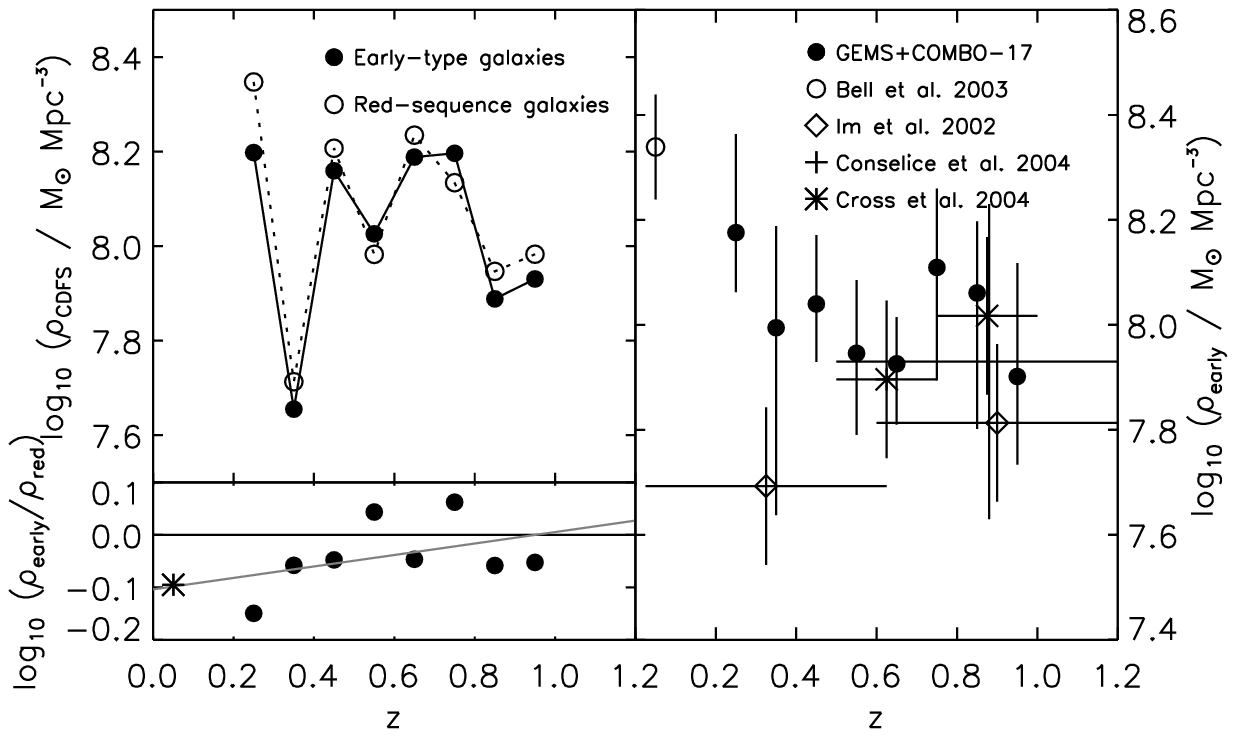}
  \caption{The evolution of the stellar mass density in 
  early-type galaxies.
  Stellar masses assume a \protect\cite{kroupa01} IMF and 
  $H_0 = 70$\,km\,s$^{-1}$\,Mpc$^{-1}$.
  The left-hand panel shows the stellar mass densities from GEMS
  only; solid circles denote early-type galaxies and open circles
  red-sequence galaxies.  The ratio of stellar mass in early-type
  galaxies to the red sequence is shown in the lower left panel;
  the asterisk is the local value taken from \protect\cite{bell03},
  and the grey line shows a linear fit to the GEMS data only (RMS
  $\sim 15$\%).  The right-hand panel shows the resulting 
  corrected early-type galaxy stellar mass density evolution, 
  where again the local data are taken from \protect\cite{bell03},
  the solid circles denote the GEMS$+$COMBO-17 result, 
  the diamonds show results from \protect\cite{im02}, 
  the naked error bars show results from \protect\cite{conselice04}, 
  and the asterisks show results from \protect\cite{cross04}.
  }\label{fig:early}
  \end{center}
\end{figure}

\cite{im02} present a thorough study of the luminosity density
evolution of E/S0 galaxies 
from the DEEP Groth Strip Survey, defined using bulge--disk 
decompositions and placing a limit on residual substructure in 
the model-subtracted images, supplemented with 
HST $V$ and $I$-band imaging for 118 square arcminutes.
A final sample of 145 E/S0 galaxies with $0.05 < z < 1.2$ are 
isolated.  The fit of average early-type galaxy stellar
M/L as a function of redshift from COMBO-17/GEMS; 
$\log_{10} {\rm M/L}_B = 0.34 - 0.27z$, 
was used to transform the published
$B$-band luminosity densities into stellar mass densities
for the purposes of Fig.\ \ref{fig:early}\footnote{Again, a
\cite{kroupa01} IMF was used.}.  \cite{cross04} 
present rest-frame $B$-band luminosity functions for
visually classified E/S0 galaxies in 5 ACS fields:
the $B$-band luminosity densities were transformed
into stellar mass in exactly the same way as 
for \cite{im02}.  \cite{conselice04} presented stellar 
mass densities directly, and these are reproduced on Fig.\ \ref{fig:early},
after accounting for our use of a \cite{kroupa01} IMF.

The preliminary GEMS/COMBO-17 early-type galaxy data points depict
the evolution of total stellar mass in galaxies with S\'ersic
indices $n > 2.5$.  
In GEMS, \cite{bellgems} found that galaxies with
$n > 2.5$ included $\sim 80$\% of the visually-classified
E/S0/Sa galaxy population at $z \sim 0.7$, with $\sim 20$\% contamination 
from later galaxy types (recall, high S\'ersic indices 
indicate concentrated light profiles).  Here, in this very preliminary
exploration of the issue, a 
$n = 2.5$ cut is adopted irrespective of redshift, ignoring the important
issue of morphological $k$-correction.  To recap, stellar
masses are calculated using color-dependent stellar 
M/Ls from \cite{ml}, again extrapolating to total using a faint-end
slope $\alpha = -0.6$.

The resulting $n > 2.5$ stellar mass evolution for the $30' \times 30'$ GEMS
field is shown in the left panel of 
Fig.\ \ref{fig:early}.  One can clearly see strong variation
in the stellar mass density of early-type galaxies, resulting from 
large scale structure along the line of sight.  From such data, it
is clearly not possible to place any but the most rudimentary and uninteresting
limits on the evolution of early-type galaxies over the last 9 Gyr.  
Yet, noting that the bulk of early-type galaxies are
in the red sequence at $z \sim 0$ 
(e.g., \cite[Strateva et al.\ 2001]{strateva01}) and
at $z \sim 0.7$ (e.g., \cite[Bell et al.\ 2004a]{bellgems}), and
that the stellar mass density of red-sequence galaxies undergo
very similar fluctuations, it becomes interesting to 
ask if the ratio of stellar mass in red-sequence galaxies to
early types varies more smoothly with redshift.  This is plotted in
the lower left panel of Fig.\ \ref{fig:early}.  There is a weak trend
in early-type galaxy to red-sequence galaxy mass
density, caused by an increasingly important
population of blue galaxies with $n > 2.5$ towards
higher and higher redshift (see also 
\cite[Cross et al.\ 2004]{cross04}, who discuss 
this issue in much more depth).  Importantly, however, there
is only a $\sim 15$\% scatter around this trend despite the nearly
order of magnitude variation in galaxy density, 
arguing against strong environmental 
dependence in the early type to red galaxy ratio.

This relatively slow variation in early-type to red-sequence
ratio (modeled using a simple linear fit for the purposes of 
this paper) is used to convert COMBO-17's stellar mass
evolution in red galaxies from Fig.\ \ref{fig:red}, which is much
less sensitive to large scale structure, into the evolution of 
stellar mass density in morphologically early-type galaxies, 
which is shown in the right panel of Fig.\ \ref{fig:early}.
It is clear that there is a strongly increasing stellar mass
density in morphologically early-type galaxies since 
$z \sim 1.2$\footnote{\protect\cite{im02} report a low stellar
mass density in early-type galaxies at $0.05 < z < 0.6$.  We would
attribute this deficiency in stellar mass to incompleteness
at bright apparent magnitudes, leading to a deficit of nearby
E/S0s with large stellar masses (as argued by Im et al.\ themselves), 
and perhaps to small number statistics and cosmic variance (as the volume
probed by this study at $z \lsim 0.5$ is rather small).  
Further work is required 
to explore further this discrepancy. }.
While there are recent indications that lower-luminosity
early-type galaxies are largely absent at $z \gsim 0.8$ (e.g.,
\cite[Kodama et al.\ 2004]{kodama04}, 
\cite[de Lucia et al.\ 2004]{delucia04}), the 
total stellar mass density is strongly dominated by $\sim L^*$ 
galaxies, and there is no room to avoid the conclusion that there has been 
a substantial build-up in the total number of 
$\sim L^*$ early-type galaxies since $z \sim 1.2$. 
Like the results presented in sections \ref{sec:close} and \ref{sec:mergers},
these results suggest an important role for $z \lsim 1$ 
galaxy mergers in shaping the present-day galaxy 
population\footnote{Disk re-accretion and fading work in opposite
directions in this context; disk accretion following a major
merger takes galaxies away from the early-type class, whereas
fading of disks which formed at earlier times will increase
the relative prominence of spheroidal bulge components and add
galaxies to the early-type class.  A thorough 
investigation of these issues, involving bulge--disk decompositions
of rest-frame $B$-band images of the GEMS sample and drawing on 
galaxy evolution models to build intuition of the importance
and effects of different physical processes, is in preparation
(H\"au{\ss}ler et al., in prep.).}.

\subsection{The importance of galaxy mergers since $z \sim 1$}

Three independent methods have been brought to bear on this
problem --- the evolution of close galaxy pairs, the evolution 
of morphologically-disturbed galaxies, and the evolution 
of early-type galaxies as a proxy for plausible merger remnants.
All three methods suffer from important systematic and 
interpretive difficulties.  Close galaxy pairs, even supplemented
with spectroscopy to isolate galaxies within $\lsim 500$\,kms$^{-1}$
of each other, remain contaminated by galaxies in the same 
local structures.  Some measures of morphological
disturbance are susceptible to this source of error
to a lesser extent, and a clear consensus on the meaning 
of the different automated and visual measures of morphological
disturbance is yet to emerge.  Early-type galaxies will
be the result of only a subset of galaxy mergers and interactions,
and the role of disk re-accretion and fading in driving early-type
galaxy evolution is frustratingly unclear.

Yet, despite these difficulties some broad features are clear.
Mergers between $\sim L^*$ galaxies are almost certainly more frequent
at higher redshift than at the present day, but this does not 
imply by any means that galaxy mergers are unimportant at $z \sim 1$:
indeed, it is possible that an average $\sim L^*$ galaxy 
undergoes roughly 1 merger between $z = 1$ and the present day
(\cite[Le F\`evre et al.\ 2000]{lefevre00}).  This is supported
by the substantial build-up in stellar mass in red-sequence and early-type
galaxies since $z \sim 1$.  Substantial uncertainties remain and
important questions, like the stellar mass dependence of the merger 
rate, or the fraction of dissipationless vs.\ dissipational 
mergers, are completely open.  Nonetheless, these
first, encouraging steps imply that the massive galaxy population 
is strongly affected by late galaxy 
mergers, in excellent qualitative agreement with 
our understanding of galaxy evolution in a $\Lambda$CDM Universe.

\section{Summary and Outlook}

The last decade has witnessed impressive progress in our 
understanding of galaxy formation and evolution.  Despite 
important technical and interpretive difficulties, 
the broad phenomenology of galaxy assembly has been 
described with sufficient accuracy to start constraining
models of galaxy formation and evolution.  The history of 
star formation has attacked from two complementary and largely
independent angles --- through the evolution of the cosmic-averaged
star formation rate, and through the build-up of stellar mass 
with cosmic time.  Prior to $z \sim 1$, stars formed
rapidly, and $\sim 2/3$ of the total present-day stellar
mass was formed in this short $\sim 4$\,Gyr interval.  
The epoch subsequent to $z \sim 1$ has seen
a dramatic decline in cosmic-averaged SFR by a factor of 
roughly 10; however, $\sim 1/3$ of the present-day stellar
mass was formed in this 9 Gyr interval.  These two 
diagnostics of cosmic star formation history paint a largely consistent
picture, giving confidence in its basic features.

The assembly of present-day galaxies from their progenitors
through the process of galaxy mergers 
was studied using three largely independent 
diagnostics --- the evolution of galaxy close pairs, 
the evolution of morphologically-disturbed galaxies, 
and the evolution of early-type galaxies as a plausible
major merger remnant.  The interpretive difficulties 
plaguing all three diagnostics are acute; accordingly
our understanding of the importance of major mergers
in shaping the present-day galaxy population is 
incomplete.  Yet, all three diagnostics seem to indicate
that an important fraction (dare I suggest $\gsim 1/2$?) 
of $\sim L^* \sim 3 \times 10^{10} M_{\odot}$ 
galaxies are affected by galaxy mergers
since $z \sim 1$.  This contrasts sharply with 
the form of the cosmic star formation history, 
where all of the action is essentially over by $z \sim 1$.
It is not by any means indefensible to argue that 
late mergers shape the properties of the massive galaxy population
in a way which is qualitatively consistent with our
understanding of galaxy formation and evolution in 
a $\Lambda$CDM Universe.  

Yet, it is clear that much work is to be done to 
fully characterize the physical processes driving
galaxy evolution in the epoch since reionization $z \lsim 6$.
The `shopping list' is too extensive to discuss properly,
so I will focus on three aspects which I feel are particularly
important.  

\subsection{Our increasing understanding of the local Universe}

Wide area, uniform photometric and spectroscopic
surveys, such as the Sloan Digital Sky Survey
(SDSS; \cite[York et al.\ 2000]{york00}), the 
Two Micron All Sky Survey (2MASS; \cite[Skrutskie et al.\ 1997]{skrut97}), and 
the Two degree Field Galaxy Redsift Survey 
(2dFGRS; \cite[Colless et al.\ 2001]{colless01}) 
are revolutionizing our understanding of the local galaxy
population.  These surveys have allowed one to tie down the 
$z = 0$ data point for many evolutionary studies, greatly
increasing the redshift and time-baseline leverage.
Yet, this increased leverage is often difficult
to fully apply, because local studies suffer from very different
systematics than the lookback studies, and experimental
details such as imaging depth, resolution, waveband, etc.\
are often imperfectly matched.  At this stage, few groups have grasped
the nettle of repackaging these local surveys
in a format which can be readily artificially redshifted, etc.\
to allow for uniform analysis of the distant and local control 
samples.  This will be an important feature of 
the most robust of the future works in galaxy evolution,
and will greatly increase the scope and discriminatory
power of studies of galaxy photometric, dynamical, and morphological evolution.

\subsection{Using models as interpretive tools}

Large and uniform multi-wavelength and/or spectroscopic
datasets are becoming increasingly common.  An important 
consequence of this change in the nature of the data used
to study galaxy evolution is that often the error budget
is completely dominated by systematic and interpretive
uncertainties.  In this context, models of galaxy formation 
and evolution can help to understand and limit these 
systematic uncertainties through exhaustive
analysis of realistic mock catalogs.  A crude example
of this kind of analysis is given in \S \ref{sec:close}; 
a very nice example is the 2dFGRS group catalog of \cite{eke04}.
Mock simulations of this kind of quality must become 
a more prominent part of our toolkit in order for the
kind of interpretive difficulties faced in \S \ref{sec:close}
or \S \ref{sec:early} to be successfully navigated.

\subsection{The importance of multiple large HST fields}

Large scale structure is an important and frequently neglected
source of systematic uncertainty 
(\cite[Somerville et al.\ 2004]{somerville04})\footnote{It is
interesting to note that a $\times 10$ increase in area
in a single contiguous field gives only a $\times 2$ reduction 
in cosmic variance, because the various parts of the single
field are correlated with each other.}.  Comparison of 
the left-hand and right-hand panels of 
Fig.\ \ref{fig:early} illustrates this point 
powerfully; the broad features of the evolution of early-type
galaxy stellar mass density are discernable using the 
$30' \times 30'$ Chandra Deep Field South alone, but 
correction of this result for cosmic variance using 
the other two COMBO-17 fields yields a significantly
more convincing picture.  Yet, while this kind of 
correction for cosmic variance may work when there is 
significant overlap between populations of interest
(although it is debatable how far one should push such an idea),
it is not {\it a priori} clear that rarer and/or optically-obscured
phases of galaxy evolution such as AGN or IR-luminous
mergers will be well modelled with such techniques.  
Furthermore, short-timescale
astronomical phenomena, such as AGN or galaxy mergers,
have lower number density and are potentially very strongly
clustered leading to large uncertainties from number
statistics and cosmic variance.
Yet, these phases of galaxy evolution, where
galaxies undergo important and potentially permanent transformations
in the cosmic blink of an eye, \emph{require} HST-resolution 
data in order to explore their physical drivers.

When HST could be viewed as an essentially endless resource,
a piecemeal approach was perfectly optimal: more and/or larger
HST fields could be justified on a case-by-case basis, depending 
on the science goals of interest.  Yet, faced with an 
unclear future for HST, it is not obvious that this approach 
is optimal.  HST perhaps should be thought of as a fixed lifetime
experiment, where the primary goal could become the creation of 
an archival dataset which will support 10--20 years of 
top-class research.  In the creation of such an archival dataset, 
important questions will need to be addressed: availability of 
resources will need to be balanced against number statistics 
and cosmic variance, arguably at least 2 HST passbands 
will be required to allow
some attempt at morphological $k$-correction, and deep multi-wavelength data 
will be required for study of black hole accretion and obscured
star formation, naturally
driving the fields into one of a small number of low HI and cirrus
holes (see Papovich, these proceedings).

\begin{acknowledgments}
I wish to warmly thank the GEMS and COMBO-17 collaborations ---
Marco Barden, Steven Beckwith, Andrea Borch, John Caldwell, Simon Dye,
Boris H\"au{\ss}ler, Catherine Heymans, 
Knud Jahnke, Martina Kleinheinrich, Shardha Jogee, 
Daniel McIntosh, Klaus Meisenheimer, Chien Peng, Hans-Walter Rix, 
Sebastian Sanchez, 
Rachel Somerville, Lutz Wisotzki, and last but by no means least 
Christian Wolf --- for their permission to present some
GEMS and COMBO-17 results before their publication, for 
useful discussions, and 
for their friendship and collaboration.  It is a joy to be part
of these teams.  I wish also to thank Eelco van Kampen and his
collaborators for their efforts to construct mock COMBO-17 catalogs,
for their permission to share results from these catalogs
in this article, and for useful comments.
Chris Conselice, Emmanuele Daddi, Sadegh Kochfar, and Casey Papovich 
are thanked for useful and thought-provoking discussions on some of the topics
discussed in this review, and Chris Conselice and Shardha Jogee are thanked 
for constructive comments on the first version of this review.
This work is supported by the European
Community's Human Potential Program under contract
HPRN-CT-2002-00316, SISCO.
\end{acknowledgments}

%\appendix


\begin{thebibliography}{}

  \bibitem[Arimoto \& Yoshii (1987)]{arimoto87}
	\textsc{Arimoto, N., \& Yoshii, Y.} 1987  Chemical and photometric 
	properties of a galactic wind model for elliptical galaxies
	\textsc{A\&A} \textbf{173}, 23--38.

  \bibitem[Arp (1966)]{arp66}
	\textsc{Arp, H.} 1966   Atlas of Peculiar Galaxies
	\textit{ApJS} \textbf{14}, 1--20.

  \bibitem[Baldry et al.\ (2004)]{baldry04}
	\textsc{Baldry, I.\ K., Glazebrook, K., Brinkmann, J., 
	Ivezic, Z., Lupton, R.\ H., Nichol, R.\ C., \& Szalay, A.\ S.}
	2004  Quantifying the Bimodal Color--Magnitude Distribution of 
	Galaxies  \textit{ApJ} \textbf{600}, 681--694.

  \bibitem[Barnes \& Hernquist (1996)]{barnes96}
	\textsc{Barnes, J.\ E., Hernquist, L.} 1996
	Transformations of Galaxies II.\ Gasdynamics in Merging Disk Galaxies
	\textit{ApJ} \textbf{471}, 115--142.

  \bibitem[Baugh, Cole, \& Frenk (1996)]{baugh96}
	\textsc{Baugh, C.\ M., Cole, S., \& Frenk, C.\ S.} 1996
	Evolution of the Hubble sequence in hierarchical 
	models for galaxy formation
	\textit{MNRAS} \textbf{283}, 1361--1378.

  \bibitem[Bell (2002)]{bell02}
	\textsc{Bell, E.\ F.} 2002 Dust-induced Systematic Errors 
	in Ultraviolet-derived Star Formation Rates
	\textit{ApJ} \textbf{577}, 150--154.

  \bibitem[Bell (2003)]{bell03}
	\textsc{Bell, E.\ F.} 2003 Estimating Star Formation Rates
	from Far-infrared and Radio Luminosities: The Origin of 
	the Radio--Far-infrared Correlation
	\textit{ApJ} \textbf{586}, 794--813.

  \bibitem[Bell \& de Jong (2001)]{ml}
	\textsc{Bell, E.\ F., \& de Jong, R.\ S.} 2001 
	Stellar Mass-to-Light Ratios and the Tully--Fisher Relation
	\textit{ApJ} \textbf{550}, 212--229.

  \bibitem[Bell et al.\ (2002)]{bellhii}
	\textsc{Bell, E.\ F., Gordon, K.\ D., Kennicutt, Jr., R.\ C., \&
	Zaritsky, D.} 2002 The Effects of Dust in Simple Environments: 
	Large Magellanic Cloud \textsc{Hii} Regions
	\textit{ApJ} \textbf{565}, 994--1010.

  \bibitem[Bell et al.\ (2003)]{bellmf}
	\textsc{Bell, E.\ F., McIntosh, D.\ H., Katz, N., \& Weinberg, M.\ D.} 
	2003 The Optical and Near-IR Properties of Galaxies: I.\ Luminosity
	and Stellar Mass Functions
	\textit{ApJS} \textbf{149}, 289--312.

  \bibitem[Bell et al.\ (2004a)]{bellgems}
	\textsc{Bell, E.\ F., et al.} 2004a
	GEMS Imaging of Red Sequence Galaxies at $z \sim 0.7$: Dusty or Old?
	\textit{ApJ} \textbf{600}, L11--14.

  \bibitem[Bell et al.\ (2004b)]{bell04}
	\textsc{Bell, E.\ F., et al.} 2004b
	Nearly 5000 Distant Early-Type Galaxies in COMBO-17: 
	A Red Sequence and Its Evolution since $z \sim 1$
	\textit{ApJ} \textbf{608}, 752--767.

  \bibitem[Bender (1988)]{bender88}
	\textsc{Bender, R.} 1988  Rotating and counter-rotating cores
	in elliptical galaxies  \textit{A\&A} \textbf{202}, L5--8.

  \bibitem[Bendo \& Barnes (2000)]{bendo00}
	\textsc{Bendo, G.\ J., \& Barnes, J.\ E.} 2000 
	The line-of-sight velocity distributions of simulated merger remnants
	\textit{MNRAS} \textbf{316}, 315--325.

  \bibitem[Benson et al.\ (2002)]{benson02}
	\textsc{Benson, A.\ J., Lacey, C.\ G., Baugh, C.\ M., Cole, S., 
	\& Frenk, C.\ S.} 2002  The effects of photoionization on 
	galaxy formation --- I.\ Model and results at $z=0$
	\textit{MNRAS} \textbf{333}, 156--176.	

  \bibitem[Bower \& Balogh (2004)]{bower04}
	\textsc{Bower, R.\ G., \& Balogh, M.\ L.} 2004
	The Difference Between Clusters and Groups: A Journey from 
	Cluster Cores to Their Outskirts and Beyond
	\textit{Clusters of Galaxies: Probes of Cosmological Structure and Galaxy Evolution} (eds. J.\ S. Mulchaey, A. Dressler \& A. Oemler), p.\ 326
	Cambridge University Press

  \bibitem[Bower, Lucey, \& Ellis (1992)]{ble}
	\textsc{Bower, R.\ G., Lucey, J.\ R., \& Ellis, R.\ S.} 1992
	Precision Photometry of Early Type Galaxies in the Coma and 
	Virgo Clusters: a Test of the Universality of the Colour--magnitude
	Relation -- II.\ Analysis
	\textit{MNRAS} \textbf{254}, 601--603.

  \bibitem[Brinchmann \& Ellis (2000)]{brinchmann00}
	\textsc{Brinchmann, J., \& Ellis, R.\ S.} 2000
	The Mass Assembly and Star Formation Characteristics of 
	Field Galaxies of Known Morphology
	\textit{ApJ} \textbf{536}, L77--80.

  \bibitem[Bressan, Silva, \& Granato (2002)]{bressan02}
	\textsc{Bressan, A., Silva, L., \& Granato, G.\ L.} 2002
	Far infrared and radio emission in dusty starburst galaxies
	\textit{A\&A} \textbf{392}, 377--391.

  \bibitem[Bruzual \& Charlot (2003)]{bc03}
	\textsc{Bruzual, G., \& Charlot, S.} 2003
	Stellar population synthesis at the resolution of 2003
	\textit{MNRAS} \textbf{344}, 1000--1028.

  \bibitem[Bundy et al.(2004)]{bundy04}
	\textsc{Bundy, K., Fukugita, M., Ellis, R.\ S., Kodama, T., 
	\& Conselice, C.\ J.} 2004  A Slow Merger History 
	of Field Galaxies since $z \sim 1$     \textit{ApJ}
	\textbf{601}, L123--126.

  \bibitem[Calzetti (2001)]{calzetti01}
	\textsc{Calzetti, D.} 2001
	The Dust Opacity of Star-forming Galaxies
	\textit{PASP} \textbf{113}, 1449--1485.

  \bibitem[Calzetti, Kinney, \& Storchi-Bergmann (1994)]{calzetti94}
	\textsc{Calzetti, D., Kinney, A.\ L., \& Storchi-Bergmann, T.} 1994
	Dust extinction of the stellar continua in starburst galaxies: 
	The ultraviolet and optical extinction law
	\textit{ApJ} \textbf{429}, 582--601.

  \bibitem[Carlberg, Pritchet, \& Infante (1994)]{carlberg94}
	\textsc{Carlberg, R.\ G., Pritchet, C.\ J., \& Infante, L.}
	1994  A survey of faint galaxy pairs
	\textit{ApJ} \textbf{435}, 540--547.

  \bibitem[Chen et al.\ (2003)]{chen03}
	\textsc{Chen, H.-W., et al.} 2003  The Las Campanas Infrared Survey 
	IV.\ The Photometric Redshift Survey and the 
	Rest-Frame R-Band Galaxy Luminosity Function at $0.5 \le z \le 1.5$
	\textit{ApJ} \textbf{586}, 745--764.

  \bibitem[Cimatti et al.\ (2002)]{cimatti02}
	\textsc{Cimatti, A., et al.} 2002   The K20 survey -- I.\ 
	Disentangling old and dusty star-forming galaxies in the ERO population
	\textit{A\&A} \textbf{381}, L68--72.

  \bibitem[Cole et al.\ (2000)]{cole00}
	\textsc{Cole, S., Lacey, C.\ G., Baugh, C.\ M., Frenk, C.\ S.} 2000
	Hierarchical galaxy formation
	\textit{MNRAS} \textbf{319}, 168--204.

  \bibitem[Cole et al.\ (2001)]{cole01}
	\textsc{Cole, S., et al.} 2001
	The 2dF galaxy redshift survey: near-infrared 
	galaxy luminosity functions
	\textit{MNRAS} \textbf{326}, 255-273.

  \bibitem[Colless et al.\ (2001)]{colless01}
	\textsc{Colless, M., et al.} 2001  The 2dF Galaxy Redshift Survey: 
	spectra and redshifts \textit{MNRAS} \textbf{328}, 1039--1063.

  \bibitem[Condon (1992)]{condon92}
	\textsc{Condon, J.\ J.} 1992 Radio emission from normal galaxies
	\textit{ARA\&A} \textbf{30}, 575--611.

  \bibitem[Conselice et al.\ (2000)]{con00}
	\textsc{Conselice, C.\ J., Bershady, M.\ A., Jangren, A.} 2000
	The Asymmetry of Galaxies: Physical Morphology for 
	Nearby and High-Redshift Galaxies
	\textit{ApJ} \textbf{529}, 886--910.

  \bibitem[Conselice et al.\ (2003)]{conselice03}
	\textsc{Conselice, C.\ J., Bershady, M.\ A., Dickinson, M., \&
	Papovich, C.} 2003  A Direct Measurement of Major Galaxy 
	Mergers at $z \lsim 3$
	\textit{AJ} \textbf{126}, 1183--1207.

  \bibitem[Conselice et al.\ (2004)]{conselice04}
	\textsc{Conselice, C.\ J., Blackburne, J.\ A., \&
	Papovich, C.} 2004  The luminosity, stellar mass, and number
	density evolution of field galaxies of known morphology
	from $z = 0.5-3$
	submitted to the \textit{ApJ} (astro-ph/0405001).

  \bibitem[Cross et al.\ (2004)]{cross04}
	\textsc{Cross, N.\ J.\ C., et al.} 2004  
	The luminosity function of early-type field galaxies at
	$z \sim 0.75$  \textit{AJ}, in press (astro-ph/0407644).

  \bibitem[Davies et al.\ (1983)]{davies83}
	\textsc{Davies, R.\ L., Efstathiou, G., Fall, S.\ M., Illingworth, G.,
	Schechter, P.\ L.} 1983  The kinematic properties of faint elliptical
	galaxies \textit{ApJ} \textbf{266}, 41--57.

  \bibitem[de Lucia et al.\ (2004)]{delucia04}
	\textsc{de Lucia, G., et al.} 2004  The Buildup of the 
	Red Sequence in Galaxy Clusters since $z \sim 0.8$
	\textit{ApJ} \textbf{610}, L77--80.

  \bibitem[Dickinson et al.\ (2003)]{dickinson03}
	\textsc{Dickinson, M.\ E., Papovich, C., Ferguson, H.\ C., \&
	Budav\'ari, T.} 2003
        The Evolution of the Global Stellar Mass Density at $0 < z < 3$
	\textit{ApJ} \textbf{587}, 25--40.

  \bibitem[Drory et al.\ (2004)]{drory04}
	\textsc{Drory, N., Bender, R., Feulner, G., Hopp, U., Maraston, C.
	Snigula, J., \& Hill, G.\ J.} 2004
	The Munich Near-Infrared Cluster Survey (MUNICS) VI.\ 
	The Stellar Masses of $K$-Band-selected Field Galaxies to $z \sim 1.2$
	\textit{ApJ} \textbf{608}, 742--751.

  \bibitem[Eggen, Lynden-Bell, \& Sandage (1962)]{eggen62}
	\textsc{Eggen, O.\ J., Lynden-Bell, D., Sandage, A.\ R.} 1962
	Evidence from the motions of old stars that the Galaxy collapsed
	\textit{ApJ} \textbf{136}, 748--766.

  \bibitem[Eke et al.\ (2004)]{eke04}
	\textsc{Eke, V.\ R., et al.} 2004   Galaxy groups in the 2dFGRS: 
	the group-finding algorithm and the 2PIGG catalogue
	\textit{MNRAS} \textbf{348}, 866--878.

  \bibitem[Ferrarese et al.\ (2001)]{ferrarese01}
	\textsc{Ferrarese, L., Pogge, R.\ W., Peterson, B.\ M., 
	Merritt, D., Wandel, A., Joseph, C.\ L.} 2001
	Supermassive Black Holes in Active Galactic Nuclei -- I.\ 
	The Consistency of Black Hole Masses in Quiescent and Active Galaxies
	\textit{ApJ} \textbf{555}, L79--82.

  \bibitem[Fioc \& Rocca-Volmerange (1997)]{pegase}
	\textsc{Fioc, M., \& Rocca-Volmerange, B.} 1997
        PEGASE: a UV to NIR spectral evolution model of galaxies. 
	Application to the calibration of bright galaxy counts.
	\textit{A\&A} \textbf{326}, 950--962.

  \bibitem[Flores et al.\ (1999)]{flores99}
	\textsc{Flores, H., et al.} 1999 
        15$\mu$m ISO Observations of the 1415+52 CFRS Field: The Cosmic SFR 
	as Derived from Deep UV, Optical, Mid-IR, and Radio Photometry
	\textit{ApJ} \textbf{517}, 148--167.

  \bibitem[Fontana et al.\ (2003)]{fontana03}
	\textsc{Fontana, A., et al.} 2003 
        The Assembly of Massive Galaxies from Near-Infrared 
	Observations of the Hubble Deep Field South
	\textit{ApJ} \textbf{594}, L9--12.

  \bibitem[Freedman et al.\ (2001)]{freedman01}
	\textsc{Freedman, W.\ L., et al.} 2001
	Final Results from the Hubble Space Telescope 
	Key Project to Measure the Hubble Constant
 	\textit{ApJ} \textbf{553}, 47--72.

  \bibitem[Genzel et al.\ (2001)]{genzel01}
	\textsc{Genzel, R., Tacconi, L.\ J., Rigopoulou, D., 
	Lutz, D., Tecza, M.} 2001  Ultraluminous Infrared
	Mergers: Elliptical Galaxies in Formation?
	\textit{ApJ} \textbf{563}, 527--545.

  \bibitem[Glazebrook et al.\ (2004)]{glaze04}
	\textsc{Glazebrook, K., et al.} 2004  A high abundance of massive
	galaxies 3--6 billion years after the Big Bang  
	\textit{Nature} \textbf{430}, 181--184.

  \bibitem[Goldader et al.\ (2002)]{goldader02}
	\textsc{Goldader, J.\ D., et al.} 2002
	Far-Infrared Galaxies in the Far-Ultraviolet
	\textit{ApJ} \textbf{568}, 651--678.

  \bibitem[Gordon et al.\ (2004)]{gordon04}
	\textsc{Gordon, K.\ D., et al.} 2004
	Spatially Resolved Ultraviolet, H$\alpha$, Infrared, 
	and Radio Star Formation in M81
 	\textit{ApJS} in press. (astro-ph/0406064)

  \bibitem[Haarsma et al.\ (2000)]{haarsma00}
	\textsc{Haarsma, D.\ B., Partridge, R.\ B., Windhorst, R.\ A., \&
	Richards, E.\ A.} 2000 Faint Radio Sources and Star Formation History
	\textit{ApJ} \textbf{544}, 641--658.

  \bibitem[Haehnelt (2004)]{haehnelt04}
	\textsc{Haehnelt, M.\ G.} 2004  Joint Formation of Supermassive
	Black Holes and Galaxies.  In \emph{Coevolution of Black Holes 
	and Galaxies} (ed.\ L.\ C.\ Ho),  p. 406.  Cambridge University Press.

  \bibitem[Hogg (2001)]{hogg01}
	\textsc{Hogg, D.\ W.} 2001 A meta-analysis of cosmic 
	star-formation history \textit{submitted to PASP} (astro-ph/0105280)

  \bibitem[Hopkins (2004)]{hopkins04}
	\textsc{Hopkins, A.\ M.} 2004
	On the Evolution of Star Forming Galaxies
 	\textit{ApJ} in press. (astro-ph/0407170)

  \bibitem[Im et al.(2002)]{im02}
	\textsc{Im, M., et al.} 2002  The DEEP Groth Strip Survey X.\ 
	Number Density and Luminosity Function of Field E/S0 Galaxies at $z<1$
	\textit{ApJ} \textbf{571}, 136--171.

  \bibitem[Kauffmann et al.\ (2003)]{kauffmann03}
	\textsc{Kauffmann, G., et al.} 2003
	Stellar masses and star formation histories for 10$^5$ 
	galaxies from the Sloan Digital Sky Survey
 	\textit{MNRAS} \textbf{341}, 33--53.

  \bibitem[Kelson et al.\ (2001)]{kelson01}
	\textsc{Kelson, D.\ D., Illingworth, G.\ D., Franx, M., \&
	van Dokkum, P.\ G.} 2001 The Evolution of Balmer 
	Absorption-Line Strengths in E/S0 Galaxies from $z=0$ to $z=0.83$
	\textit{ApJ} \textbf{522}, L17--21.

  \bibitem[Kennicutt (1983)]{kennicutt83}
	\textsc{Kennicutt, Jr., R.\ C.} 1983
	The rate of star formation in normal disk galaxies
 	\textit{ApJ} \textbf{272}, 54--67.

  \bibitem[Kennicutt (1998)]{ke98}
	\textsc{Kennicutt, Jr., R.\ C.} 1998
	Star Formation in Galaxies Along the Hubble Sequence
 	\textit{ARA\&A} \textbf{36}, 189--232.

  \bibitem[Kodama et al.\ (2004)]{kodama04}
	\textsc{Kodama, T., et al.} 2004  Down-sizing in galaxy 
	formation at $z \sim 1$ in the Subaru/XMM-Newton Deep Survey (SXDS)
	\textit{MNRAS} \textbf{350}, 1005--1014.

  \bibitem[Kong et al.\ (2004)]{kong04}
	\textsc{Kong, X., Charlot, S., Brinchmann, J., Fall, S.\ M.} 2004
	Star formation history and dust content of galaxies 
	drawn from ultraviolet surveys
 	\textit{MNRAS} \textbf{349}, 769--778.

  \bibitem[Kroupa (2001)]{kroupa01}
	\textsc{Kroupa, P.} 2001
	On the variation of the initial mass function
 	\textit{MNRAS} \textbf{322}, 231--246.

  \bibitem[Larson (1974)]{larson74}
	\textsc{Larson, R.\ B.} 1974 Dynamical models for the 
	formation and evolution of spherical galaxies
	\textit{MNRAS} \textbf{166}, 585--616.

  \bibitem[Le F\`evre et al.\ (2000)]{lefevre00}
	\textsc{Le F\`evre, O., et al.} 2000   Hubble Space Telescope 
	imaging of the CFRS and LDSS redshift surveys -- IV.\ 
	Influence of mergers in the evolution of faint field galaxies from 
	$z \sim 1$  \textit{MNRAS} \textbf{311}, 565--575.

  \bibitem[Lilly et al.\ (1995)]{lilly95}
	\textsc{Lilly, S.\ J., Tresse, L., Hammer, F., Crampton, D., 
	\& Le F\`evre, O., } 1995   The Canada-France Redshift Survey: VI.\ 
	Evolution of the Galaxy Luminosity Function to $z \sim 1$
 	\textit{ApJ} \textbf{455}, 108--124.

  \bibitem[Lilly et al.\ (1996)]{lilly96}
	\textsc{Lilly, S.\ J., Le F\`evre, O., Hammer, F., \& Crampton, D.}
	1996 The Canada-France Redshift Survey: 
	The Luminosity Density and Star Formation History of 
	the Universe to $z \sim 1$
 	\textit{ApJ} \textbf{460}, L1-4	

  \bibitem[Lisenfeld, V\"olk, \& Xu (1996)]{lisenfeld96}
	\textsc{Lisenfeld, U., V\"olk, H.\ J., \& Xu, C.} 1996
	The FIR/radio correlation in starburst galaxies: 
	constraints on starburst models
 	\textit{A\&A} \textbf{314}, 745--753.

  \bibitem[Lotz, Primack, \& Madau (2004)]{lotz04}
	\textsc{Lotz, J.\ M., Primack, J., \& Madau, P.} 2004
	A New Nonparametric Approach to Galaxy Morphological Classification
	\textit{AJ} \textbf{128}, 163--182.

  \bibitem[Madau et al.\ (1996)]{madau96}
	\textsc{Madau, P., Ferguson, H.\ C., Dickinson, M.\ E., 
	Giavalisco, M., Steidel, C.\ C., \& Fruchter, A.} 1996
	High-redshift galaxies in the Hubble Deep Field: 
	colour selection and star formation history to $z \sim 4$
	\textit{MNRAS} \textbf{283}, 1388--1404.

  \bibitem[Meza et al.\ (2003)]{meza03}
	\textsc{Meza, A., Navarro, J.\ F., Steinmetz, M., \& Eke, V.\ R.}
	2003 Simulations of Galaxy Formation in a $\Lambda$CDM Universe 
	III.\ The Dissipative Formation of an Elliptical Galaxy
	\textit{ApJ} \textbf{590}, 619--635.

  \bibitem[Moustakas et al.\ (2004)]{moustakas04}
	\textsc{Moustakas, L.\ A., et al.} 2004
	Morphologies and Spectral Energy Distributions of 
	Extremely Red Galaxies in the GOODS-South Field
	\textit{ApJ} \textbf{600}, L131--134.

  \bibitem[Naab \& Burkert (2003)]{naab03}
	\textsc{Naab, T., \& Burkert, A.} 2003  Statistical Properties 
	of Collisionless Equal- and Unequal-Mass Merger Remnants of 
	Disk Galaxies  \textit{ApJ} \textbf{597}, 893--906.

  \bibitem[Niklas \& Beck (1997)]{niklas97}
	\textsc{Niklas, S., \& Beck, R.} 1997
	A new approach to the radio--far infrared 
	correlation for non-calorimeter galaxies
 	\textit{A\&A} \textbf{320}, 54--64.

  \bibitem[Papovich, Dickinson, \& Ferguson (2001)]{papovich01}
	\textsc{Papovich, C., Dickinson, M.\ E., \& Ferguson, H.\ C.} 2001
	The Stellar Populations and Evolution of Lyman Break Galaxies
 	\textit{ApJ} \textbf{559}, 620--653.

  \bibitem[Patton et al.\ (2000)]{patton00}
	\textsc{Patton, D.\ R., et al.} 2000  New Techniques for Relating
	Dynamically Close Galaxy Pairs to Merger and Accretion Rates: 
	Application to the Second Southern Sky Redshift Survey
	\textit{ApJ} \textbf{536}, 153--172.

  \bibitem[Patton et al.\ (2002)]{patton02}
	\textsc{Patton, D.\ R., et al.} 2002  Dynamically Close Galaxy
	Pairs and Merger Rate Evolution in the CNOC2 Redshift Survey
	\textit{ApJ} \textbf{565}, 208--222.

  \bibitem[Peebles (1980)]{peebles80}
	\textsc{Peebles, P.\ J.\ E.} 1980  \emph{The large-scale 
	structure of the universe}.  Princeton, N.\ J.: Princeton
	University Press. 

  \bibitem[Riess et al.\ (2004)]{riess04}
	\textsc{Riess, A.\ G., et al.} 2004 Type Ia Supernova Discoveries at 
	$z > 1$ from the Hubble Space Telescope: Evidence for Past 
	Deceleration and Constraints on Dark Energy Evolution
	\textit{ApJ} \textbf{607}, 665--687.

  \bibitem[Rix et al.\ (2004)]{rix04}
	\textsc{Rix, H.-W., et al.} 2004  GEMS: Galaxy Evolution from 
	Morphology and SEDs  \textit{ApJS} \textbf{152}, 163--173.

  \bibitem[Rudnick et al.\ (2003)]{rudnick03}
	\textsc{Rudnick, G., et al.} 2003 The Rest-Frame Optical Luminosity
	Density, Color, and Stellar Mass Density of the Universe 
	from $z = 0$ to $z = 3$
	\textit{ApJ} \textbf{599}, 847--864.

  \bibitem[Sandage \& Visvanathan (1978)]{sandage78}
	\textsc{Sandage, A., \& Visvanathan, N.} 1978
	The color-absolute magnitude relation for E and S0 galaxies. 
	II -- New colors, magnitudes, and types for 405 galaxies
	\textit{ApJ} \textbf{223}, 707--729.

  \bibitem[Schweizer \& Seitzer (1992)]{schweizer92}
	\textsc{Schweizer, F., \& Seitzer, P.} 1992
	Correlations between UBV colors and fine structure in E and 
	S0 galaxies --- A first attempt at dating ancient merger events
	\textit{AJ} \textbf{104}, 1039--1067.

  \bibitem[Seljak et al. (2004)]{seljak04}
	\textsc{Seljak, U., et al.} 2004  Cosmological parameter analysis 
	including SDSS Ly-alpha forest and galaxy bias: constraints on 
	the primordial spectrum of fluctuations, neutrino mass, and dark energy
	\textit{submitted to PRD} (astro-ph/0407372).

  \bibitem[S\'{e}rsic (1968)]{sersic68}
        \textsc{S\'{e}rsic, J. L.} 1968 \emph{Atlas de Galaxias Australes}.
	 Cordoba: Observatorio Astronomico.

  \bibitem[Skrutskie et al.\ (1997)]{skrut97}
	\textsc{Skrutskie, M.\ F., et al.} 1997  The Two Micron 
	All Sky Survey (2MASS): Overview and Status.
	in \emph{The Impact of Large Scale Near-IR Sky Surveys} 
	(eds. F.\ Garzon et al.) p. 25.  Dordrecht: 
	Kluwer Academic Publishing Company

  \bibitem[Somerville et al.\ (2004)]{somerville04}
	\textsc{Somerville, R.\ S., Lee, K., Ferguson, H.\ C., 
	Gardner, J.\ P., Moustakas, L.\ A., \& Giavalisco, M.} 2004
	Cosmic Variance in the Great Observatories Origins Deep Survey
	\textit{ApJ} \textbf{600}, L171--174.

  \bibitem[Spergel et al.\ (2003)]{spergel03}
	\textsc{Spergel, D.\ N., et al.} 2003
	First-Year Wilkinson Microwave Anisotropy Probe (WMAP) 
	Observations: Determination of Cosmological Parameters
 	\textit{ApJS} \textbf{148}, 175--194.

  \bibitem[Stanway et al.\ (2004)]{stan04}
	\textsc{Stanway, E.\ R., Bunker, A.\ J., McMahon, R.\ G.,
	Ellis, R.\ S., Treu, T., \& McCarthy, P.\ J.} 2004
	Hubble Space Telescope Imaging and Keck Spectroscopy of $z \sim 6$ 
	$i$-Band Dropout Galaxies in the Advanced Camera for 
	Surveys GOODS Fields
	\textit{ApJ} \textbf{607}, 704--720.

  \bibitem[Steidel et al.\ (1999)]{steidel99}
	\textsc{Steidel, C.\ C., Adelberger, K.\ L., Giavalisco, M., 
	Dickinson, M., \& Pettini, M.} 1999
	Lyman-Break Galaxies at $z \gsim 4$ and the Evolution of 
	the Ultraviolet Luminosity Density at High Redshift
	\textit{ApJ} \textbf{519}, 1--17.

  \bibitem[Strateva et al.\ (2001)]{strateva01}
	\textsc{Strateva, I., et al.} 2001  Color Separation of 
	Galaxy Types in the Sloan Digital Sky Survey Imaging Data
	\textit{AJ} \textbf{122}, 1861--1874.

  \bibitem[Toomre \& Toomre (1972)]{toomre72}
	\textsc{Toomre, A., \& Toomre, J.} 1972  Galactic Bridges and Tails
	\textit{ApJ} \textbf{178}, 623--666.

  \bibitem[Tully et al.\ (1998)]{tully98}
     \textsc{Tully, R.\ B., Pierce, M.\ J., Huang, J.\ S., Saunders, W.,
	Verheijen, M.\ A.\ W., \& Witchalls, P.\ L.} 1998 
	Global Extinction in Spiral Galaxies
	\textit{AJ} \textbf{115}, 2264--2272.

  \bibitem[van Kampen, Jimenez, \& Peacock (1999)]{vankampen99}
     \textsc{van Kampen, E., Jimenez, R., \& Peacock, J.\ A.} 1999
	Overmerging and mass-to-light ratios 
	in phenomenological galaxy formation models
	\textit{MNRAS} \textbf{310}, 43--56.

  \bibitem[White \& Frenk (1991)]{white91}
     \textsc{White, S.\ D.\ M., \& Frenk, C.\ S.} 1991 
	Galaxy formation through hierarchical clustering 
	\textit{ApJ} \textbf{379}, 52--79.

  \bibitem[White \& Rees (1978)]{white78}
     \textsc{White, S.\ D.\ M., \& Rees, M.\ J.} 1978 Core condensation 
	in heavy halos --- A two-stage theory for galaxy formation and 
	clustering \textit{MNRAS} \textbf{183}, 341--358.

  \bibitem[Witt \& Gordon (2000)]{witt00}
	\textsc{Witt, A.\ N., \& Gordon, K.\ D.} 2000 
	Multiple Scattering in Clumpy Media II.\ Galactic Environments
	\textit{ApJ} \textbf{528}, 799--816.

  \bibitem[Wolf et al.\ (2003)]{wolf03}
	\textsc{Wolf, C., Meisenheimer, K., Rix, H.-W., Borch, A., Dye, S.,
	\& Kleinheinrich, M.} 2003
	The COMBO-17 survey: Evolution of the galaxy luminosity 
	function from 25 000 galaxies with $0.2< z <1.2$
	\textit{A\&A} \textbf{401}, 73--98.

  \bibitem[Worthey (1994)]{worthey94}
	\textsc{Worthey, G.} 1994 Comprehensive stellar population 
	models and the disentanglement of age and metallicity effects
	\textit{ApJS} \textbf{95}, 107--149.

  \bibitem[Wyuts et al.\ (2004)]{wyuts04}
	\textsc{Wyuts, S., van Dokkum, P.\ G., Kelson, D.\ D., 
	Franx, M., Illingworth, G.\ D.} 2004   
	The Detailed Fundamental Plane of Two High-Redshift 
	Clusters: MS 2053-04 at $z=0.58$ and MS 1054-03 at $z=0.83$
	\textit{ApJ} \textbf{605}, 677--688.

  \bibitem[Yan \& Thompson (2003)]{yan03}
	\textsc{Yan, L., \& Thompson, D.} 2003  Hubble Space 
	Telescope WFPC2 Morphologies of K-selected Extremely Red Galaxies
	\textit{ApJ} \textbf{586}, 765--779.

  \bibitem[Yee \& Ellingson (1995)]{yee95}
	\textsc{Yee, H.\ K.\ C., \& Ellingson, E.} 1995
	Statistics of close galaxy pairs from a faint-galaxy redshift survey
	\textit{ApJ} \textbf{445}, 37--45.

  \bibitem[York et al.\ (2000)]{york00}
	\textsc{York, D.\ G., et al.} 2000 The Sloan Digital Sky Survey: 
	Technical Summary \textit{AJ} \textbf{120}, 1579--1587.

  \bibitem[Zepf \& Koo (1989)]{zepf89}
	\textsc{Zepf, S.\ E., \& Koo, D.\ C.} 1989
	Close pairs of galaxies in deep sky surveys
	\textit{ApJ} \textbf{337}, 34--44.

\end{thebibliography}
\end{document}